\documentclass[a4paper,11pt]{article}
\pdfoutput=1

\usepackage{jheppub}
\usepackage{graphicx}

\newcommand{\Mp}{M_\text{Pl}}
\newcommand{\Mh}{M_h}
\newcommand{\paren}[1]{\left( #1 \right)}
\newcommand{\parenfrac}[2]{\left( \frac{#1}{#2} \right)}
\newcommand{\Higgs}{H}
\newcommand{\Ricci}{\mathcal{R}}
\newcommand{\yeff}{\lambda_\text{eff}}
\newcommand{\ymin}{\lambda_\text{eff}^\text{min}}
\newcommand{\squareparen}[1]{\left[ #1 \right]}
\newcommand{\abs}[1]{\left| #1 \right|}
\newcommand{\gp}[1]{g^{\prime #1}}

\newcommand{\GB}{G}

\title{\boldmath Higgs $\xi$-inflation for the 125--126~GeV Higgs: a two-loop analysis}

\author{Kyle Allison}
\affiliation{Rudolf Peierls Centre for Theoretical Physics, University of Oxford,\\ 1 Keble Road, Oxford OX1 3NP, United Kingdom}
\emailAdd{k.allison1@physics.ox.ac.uk}

\abstract{A non-minimal coupling $\xi$ of the Standard Model Higgs field to gravity can give rise to inflation, but large $\xi$ is required and thus leads to a violation of perturbative unitarity at $\Mp/\xi$, which is well below the inflationary scale $\Mp/\sqrt{\xi}$. We re-examine this claim for a Higgs mass in the range 125--126~GeV for which $\lambda_\text{eff}(\mu)$ runs to very small values near the Planck scale and can significantly reduce the value of $\xi$ required for inflation. Using the two-loop renormalization group equations and effective potential for Higgs $\xi$-inflation, we find that familiar inflationary solutions can have a non-minimal coupling as small as $\xi \sim 400$ without the potential developing a second minimum. We also find a new observationally allowed region of Higgs $\xi$-inflation with $\xi \sim 90$ and distinct inflationary predictions, including an observable level of the tensor-to-scalar ratio $r$.}

\begin{document}
\maketitle
\flushbottom

\section{Introduction}
A period of exponential expansion of the early universe driven by the potential energy of a scalar field --- the inflaton --- is an elegant explanation for the flatness, isotropy and homogeneity of the universe today~\cite{Sta80,Gut81,Lin82,Alb82,Lin83}. Furthermore, it provides a very plausible mechanism for generating the nearly scale invariant spectrum of primordial density fluctuations that have been imprinted on the cosmic microwave background (CMB)~\cite{WMA12,Pla13} and have grown into the large scale structure of galaxies~\cite{Lid93}. The nature of the inflaton is, however, still unknown. While a large number of inflationary models that extend the scalar degrees of freedom of the Standard Model (SM) have been proposed (see e.g.\ \cite{Lyt99,Mar13}), the possibility that the SM Higgs boson is the inflaton --- a scenario attractive for its minimality --- still remains for the model of Higgs inflation from a non-minimal coupling to gravity~\cite{Bez08}.\footnote{Other proposed models of Higgs inflation make use of special features of the SM potential that develop if the Higgs quartic coupling $\lambda$ runs to very small values.  The quasiflat SM potential considered in~\cite{Isi08}, however, predicts too large an amplitude of density fluctuations while false vacuum inflation~\cite{Mas12a,Mas12b,Mas12c} requires an additional scalar particle to achieve a graceful exit from inflation. Further possibilities, not discussed here, make use of derivative couplings of the Higgs to gravity or other non-renormalizable Higgs couplings~\cite{Ger10,Nak10,Kam11,Kam12,Her12}.}

This model of Higgs inflation, based on the work of~\cite{Sal89,Fak90,Kai95,Kom99}, makes use of a large non-minimal gravitational coupling $\xi \Higgs^\dag \Higgs \Ricci$ between the Higgs doublet $\Higgs$ and the Ricci scalar $\Ricci$.\footnote{This is the only local, gauge-invariant interaction with mass dimension four or less that can be added to the SM once gravity is included.} The effect of this coupling is to flatten the SM potential above the scale $\Mp/\sqrt{\xi}$, thereby allowing a sufficiently flat region for slow roll inflation. An analysis of the tree-level potential finds $\xi \simeq 5 \times 10^4\sqrt{\lambda}$ is required to produce the correct amplitude of primordial density fluctuations~\cite{Bez08}, which for $\Mh \simeq 125$--126~GeV~\cite{Gia13} gives $\xi \sim 2 \times 10^4$. The predictions for the spectral index and the tensor-to-scalar ratio are also well within the current 1$\sigma$ allowed regions~\cite{WMA12,Pla13}.

It has been pointed out, however, that Higgs $\xi$-inflation with the large value $\xi \sim 10^4$ suffers from a serious problem. Perturbative unitarity is violated at the scale $\Mp/\xi$, and new physics entering at $\Mp/\xi$ to restore unitarity is naively expected to contain new particles and interactions that affect the potential in an uncontrollable way~\cite{Bur09,Bar09,Bur10,Her10}.\footnote{It has been argued that the scale of perturbative unitarity violation for a large background Higgs field is higher than the small background field estimate $\Mp/\xi$ and, in particular, does not spoil the perturbative analysis of inflation~\cite{Bez09b,Bez11,Fer11}. In this case, one must make a non-trivial assumption about the new physics sector that the scale of new physics is background dependent~\cite{Ler12}. In this paper, we make the more conservative working assumption that the scale of new physics is independent of the background Higgs field and therefore must be taken to be the lowest scale of perturbative unitarity violation, $\Mp/\xi$.} The self-consistency of the model in the inflationary region $h \gtrsim \Mp/\sqrt{\xi}$ is therefore questionable. To address the issue of unitarity violation while preserving the minimality of Higgs inflation, one must make a rather strong assumption that either additional non-renormalizable Higgs interactions accompany the non-minimal coupling and restore unitarity~\cite{Ler10} or that new strong dynamics entering at $\Mp/\xi$ restores unitarity in a non-perturbative way~\cite{DeS09,Bez09b,Bez11,Bez11b}. It is unknown whether the former approach can be made consistent with quantum corrections or the effect of additional potential and Yukawa interactions~\cite{Ler12}, while it is unclear whether strong coupling in graviton exchange processes for the latter scenario can unitarize scattering cross sections without requiring new physics~\cite{Ler12}. If the latter scenario is possible, however, an approximate shift symmetry of the potential in the inflationary region $h \gtrsim \Mp/\sqrt\xi$ may keep quantum corrections to the potential under control~\cite{Bez11}.

The problem of perturbative unitarity violation in Higgs $\xi$-inflation, at least with regard to new physics entering at $\Mp/\xi$ below the inflationary scale, is perhaps not as severe as the tree-level estimate of $\xi$ suggests. A Higgs mass $M_h \simeq 125\text{--}126$~GeV is in the region that, for a top quark mass only about 2$\sigma$ below its central value, the effective Higgs quartic coupling $\yeff(\mu)$ can run to very small (positive) values near the Planck scale~\cite{Bez12,Deg12,Sha10}. The effect of small $\yeff(\mu)$ near the Planck scale is to reduce the value of $\xi$ necessary for successful inflation~\cite{Bez09a,DeS09,Bez09b} and hence push the scale of perturbative unitarity violation toward the inflationary scale. If inflation with $\xi \sim 1$ is possible for sufficiently small $\yeff(\mu)$ --- a scenario that is not yet explored --- the problem of perturbative unitarity violation occurring below the inflationary scale can be avoided.\footnote{In this case, note that although the potential during inflation $V^{1/4} \lesssim 2 \times 10^{16}$~GeV is constrained to be sub-Planckian~\cite{WMA12,Pla13}, the non-minimal coupling $\xi \Higgs^\dag \Higgs \Ricci$ with $\xi \sim 1$ is still relevant to inflation since the Higgs field $h \sim \Mp/\sqrt\xi$ is then assumed to be near the Planck scale.} Of course, an investigation of this possibility requires a proper treatment of the RG evolution and effective potential within the framework of Higgs $\xi$-inflation.

Extending the analysis of Higgs $\xi$-inflation to higher loop order is not entirely straightforward. While the renormalization group (RG) equations of the SM are perfectly adequate for describing the RG evolution below $\Mp/\xi$, there are two ambiguities in the RG evolution above $\Mp/\xi$ due to the non-minimal coupling of the Higgs. First, quantum loops involving the physical Higgs field (and not the Nambu-Goldstone bosons present in the Landau gauge) are heavily suppressed in this region~\cite{DeS09,Bez09b}. To deal with this, one can either use the chiral electroweak theory (SM with frozen radial Higgs mode) to derive the RG equations above $\Mp/\xi$~\cite{Bez09b} or one can simply use the RG equations of the SM with a suppression factor for each Higgs running in a loop~\cite{DeS09,Cla09,Ler09,Ler11}. Second, radiative corrections to the SM potential (in particular the choice of the renormalization scale $\mu(h)$) depend on whether they are computed in the Einstein or Jordan frame~\cite{Bez09a}, and it is unclear which frame should be used without knowledge of physics at the Planck scale.

In this paper, we extend the two-loop analysis of Higgs $\xi$-inflation~\cite{Bez09b,DeS09} to include the three-loop SM beta functions for the gauge couplings~\cite{Mil12} as well as the leading three-loop terms for the RG evolution of $\lambda$, the top Yukawa coupling $y_t$, and the Higgs anomalous dimension $\gamma$~\cite{Che12}. For the first time, a complete two-loop insertion of suppression factors for the {\em physical} Higgs loops, which was missing in~\cite{DeS09}, is carried out. The use of these RG equations provides a modest update to the previous analyses of Higgs $\xi$-inflation. 
The main focus of this paper, however, is to investigate the region of parameter space with $\yeff(\mu) \ll 1$ near the Planck scale that exists for the recently measured Higgs mass $M_h \simeq 125\text{--}126$~GeV and a top quark mass $M_t \sim 171$~GeV, about 2$\sigma$ below its central value.\footnote{For the top quark mass central value, the SM potential develops an instability at around $10^{11}$~GeV~\cite{Bez12,Deg12}. Since Higgs $\xi$-inflation requires the stability of the potential up to the inflationary scale $\Mp/\sqrt{\xi}$, one could interpret this result as disfavouring Higgs $\xi$-inflation at 2$\sigma$. The position advocated here is that a special region of Higgs $\xi$-inflation with $\lambda_\text{eff}(\mu) \ll 1$ exists within only 2$\sigma$ of experimental measurements.}

The paper is organized as follows. In section~\ref{sec:review}, we give a brief review of Higgs $\xi$-inflation and the tree-level analysis. In section~\ref{sec:twoloop}, the RG equations and the effective potential relevant for a two-loop analysis of Higgs $\xi$-inflation are presented. The numerical results and inflationary predictions for both the Einstein and Jordan frame renormalization prescriptions, with a particular focus on the small $\ymin$ region, are given in section~\ref{sec:results}. A summary of the results and the conclusions are given in section~\ref{sec:conclusions}.

\section{Tree-level analysis}
\label{sec:review}
Let us first briefly review Higgs $\xi$-inflation and the tree-level computation of the inflationary predictions. Although the tree-level results will differ from those in the two-loop analysis, many qualitative features of the computation will remain the same.

As an example of inflation from a non-minimally coupled scalar, Higgs $\xi$-inflation is characterized by a non-minimal gravitational coupling $\xi \Higgs^\dag \Higgs \Ricci$ between the Higgs doublet $\Higgs$ and the Ricci scalar $\Ricci$. The Lagrangian of the model is given by~\cite{Bez08}
\begin{equation}
\label{eq:lagrangian}
\mathcal{L} = \mathcal{L}_\text{SM} - \frac{M^2}{2}\Ricci - \xi \Higgs^\dag \Higgs \Ricci,
\end{equation}
where $\mathcal{L}_\text{SM}$ is the SM Lagrangian and $M$ is a mass parameter (the bare Planck mass) that can safely be identified with the present Planck mass value $\Mp = \paren{8\pi G_N}^{-1/2} \simeq 2.4 \times 10^{18}$~GeV for $\sqrt{\xi} \ll 10^{17}$. The part of \eqref{eq:lagrangian} that is relevant to inflation gives the action
\begin{equation}
\label{eq:actionJ}
S_J = \int d^4 x \sqrt{-g} \squareparen{-\frac{\Mp^2}{2}\paren{1 + \frac{2 \xi \Higgs^\dag \Higgs}{\Mp^2}}\Ricci + \paren{\partial_\mu \Higgs}^\dag \paren{\partial^\mu \Higgs} - V},
\end{equation}
where $V = \lambda\paren{\Higgs^\dag \Higgs-v^2/2}^2$ is the SM potential and the subscript $J$ denotes the Jordan frame. This is the frame in which the inflationary model is defined.

To compute the inflationary observables, it is convenient to first remove the non-minimal coupling to gravity in \eqref{eq:actionJ} by performing the conformal transformation
\begin{equation}
\label{eq:conformal}
g_{\mu \nu} \rightarrow \tilde{g}_{\mu \nu} = \Omega^2 g_{\mu \nu}, \qquad \Omega^2 = 1 + \frac{2\xi \Higgs^\dag \Higgs}{\Mp^2}.
\end{equation}
The resulting Einstein frame action is given by
\begin{equation}
\label{eq:actionEfull}
S_E = \int d^4 x \sqrt{-\tilde g} \squareparen{-\frac{\Mp^2}{2}\tilde{\Ricci} + \frac{1}{\Omega^2}\paren{\partial_\mu \Higgs}^\dag \paren{\partial^\mu \Higgs} + \frac{3\xi^2}{\Omega^4 \Mp^2}\partial_\mu(\Higgs^\dag \Higgs) \partial^\mu (\Higgs^\dag \Higgs) - \frac{V}{\Omega^4}},
\end{equation}
where $\tilde \Ricci$ is calculated with the metric $\tilde g$. The action \eqref{eq:actionEfull} simplifies greatly in the unitary gauge $\Higgs = \frac{1}{\sqrt 2} \paren{\begin{smallmatrix} 0 \\ h \end{smallmatrix}}$, which may be used for the tree-level computation, giving
\begin{equation}
\label{eq:actionEunitary}
S_E = \int d^4 x \sqrt{-\tilde g} \squareparen{-\frac{\Mp^2}{2}\tilde{\Ricci} + \frac{1}{2}\paren{\frac{\Omega^2 + 6\xi^2 h^2 / \Mp^2}{\Omega^4}}\partial_\mu h \partial^\mu h - \frac{V}{\Omega^4}},
\end{equation}
where $V = \frac{\lambda}{4}\paren{h^2-v^2}^2$ and $\Omega^2 = 1 + \xi h^2/\Mp^2$. It is also convenient to remove the non-canonical kinetic term for the Higgs field in \eqref{eq:actionEunitary} by changing to a new scalar field $\chi$, defined by
\begin{equation}
\label{eq:scalarfieldredef}
\frac{d\chi}{dh} = \sqrt{\frac{\Omega^2 + 6 \xi^2 h^2/\Mp^2}{\Omega^4}}.
\end{equation}
The Einstein frame action then takes the form
\begin{equation}
\label{eq:actionE}
S_E = \int d^4 x \sqrt{-\tilde g} \squareparen{-\frac{\Mp^2}{2}\tilde\Ricci + \frac{1}{2}\partial_\mu \chi \partial^\mu \chi - U(\chi)},
\end{equation}
where the potential is given by
\begin{equation}
\label{eq:potential}
U(\chi) = \frac{V}{\Omega^4} = \frac{\lambda(h^2-v^2)^2}{4(1+\xi h^2/\Mp^2)^2}
\end{equation}
with $h = h(\chi)$. It is the flattening of the potential $U(\chi)$ to a constant value $U_0 \equiv \lambda \Mp^4/4 \xi^2$ in the region $h \gtrsim \Mp/\sqrt \xi$ that allows slow roll inflation to occur.

The standard analysis of inflation in the slow roll approximation can be carried out for the field $\chi$ and potential $U(\chi)$. In the inflationary region $h^2 \gtrsim \Mp^2/\xi \gg v^2$, the slow roll parameters for $\xi \gg 1$ can be approximated by~\cite{Bez08b,DeS09} (see~\cite{Kai95} for exact expressions)
\begin{align}
\epsilon &= \frac{\Mp^2}{2}\parenfrac{dU/d\chi}{U}^2 \simeq \frac{4\Mp^4}{3\xi^2h^4},\label{eq:epsilon} \\
\eta &= \Mp^2\frac{d^2U/d\chi^2}{U} \simeq \frac{4\Mp^4}{3\xi^2 h^4}\paren{1-\frac{\xi h^2}{\Mp^2}},\label{eq:eta} \\
\zeta^2 &= \Mp^4 \frac{\paren{d^3U/d\chi^3}dU/d\chi}{U^2} \simeq \frac{16\Mp^6}{9\xi^3 h^6}\paren{\frac{\xi h^2}{\Mp^2}-3}.\label{eq:zeta} 
\end{align}
Slow roll ends when either $\epsilon \simeq 1$ or $\abs{\eta} \simeq 1$. For \eqref{eq:epsilon} and \eqref{eq:eta}, this occurs when $\epsilon \simeq 1$ at a field value $h_\text{end} \simeq \paren{4/3}^{1/4} \Mp/\sqrt{\xi} \simeq 1.07 \Mp/\sqrt{\xi}$. The number of e-folds of inflation as $h$ changes from $h_0$ to $h_\text{end}$ is given by~\cite{Bez09c}
\begin{equation}
\label{eq:N}
N = \int_{h_\text{end}}^{h_0} \frac{1}{\Mp^2}\frac{U}{dU/dh}\parenfrac{d\chi}{dh}^2 dh \simeq \frac{3}{4}\left[ \frac{h_0^2 - h_\text{end}^2}{\Mp^2/\xi} + \ln \paren{\frac{1+\xi h_\text{end}^2/\Mp^2}{1+\xi h_0^2/\Mp^2}} \right].
\end{equation}
The values of the parameters \eqref{eq:epsilon}--\eqref{eq:zeta} at a particular field value $h_0$, corresponding to the time at which the pivot scale $k_* \simeq 0.002$Mpc$^{-1}$ left the horizon during inflation, can be used to compare with the CMB data. This value of $h_0$ (or equivalently $N$) is a model-dependent quantity that is sensitive to the details of reheating. For Higgs $\xi$-inflation, an analysis of reheating finds that $N \simeq 59$, or equivalently $h_0 \simeq 9.14\Mp/\sqrt{\xi}$, is the value at which $k_*$ left the horizon during inflation~\cite{Bez09c,Gar09}. Using \eqref{eq:epsilon} in the WMAP9 normalization $U/\epsilon \simeq \paren{0.0274\Mp}^4$~\cite{WMA12}, the required value of $\xi$ is\footnote{The {\it Planck} 2013 normalization $U/\epsilon \simeq \paren{0.0269\Mp}^4$~\cite{Pla13} gives $\xi \simeq 18000$.}
\begin{equation}
\label{eq:xi}
\xi \simeq 48000 \sqrt \lambda = 48000 \frac{M_h}{\sqrt 2 v} \simeq 17000.
\end{equation}
The predictions for the spectral index $n_s$, the tensor-to-scalar ratio $r$, and the running of the spectral index $dn_s/d\ln k$ are given by
\begin{align}
n_s &= 1 - 6\epsilon + 2\eta \simeq 0.967, \\
r &= 16 \epsilon \simeq 0.0031, \\
\frac{dn_s}{d\ln k} &= 24 \epsilon^2 - 16 \epsilon \eta + 2 \zeta^2 \simeq 5.4 \times 10^{-4}.
\end{align}
These predictions for $n_s$ and $r$ are well within the current 1$\sigma$ allowed regions from~\cite{WMA12,Pla13}, while the prediction of $dn_s/d\ln k$ is consistent with observations at the 1--$2\sigma$ level.

\section{Two-loop analysis}
\label{sec:twoloop}

An analysis of Higgs $\xi$-inflation beyond the tree level must include both the running of the couplings and loop corrections to the (effective) potential~\cite{Bez09a,DeS09,Bez09b}. The most significant effect of these higher order corrections comes from the running of the Higgs quartic coupling $\lambda = \lambda(\mu)$. For $M_h \simeq 125$--126~GeV, it is well known that the running of $\lambda(\mu)$ --- or more specifically $\yeff(\mu)$ --- causes the SM potential to develop an instability below the Planck scale unless the top quark mass is about 2$\sigma$ below its central value~\cite{Bez12,Deg12}. Since Higgs $\xi$-inflation requires the stability of the potential up to the inflationary scale $\Mp/\sqrt \xi$, in order to realize this model of inflation one must make the moderate assumption of a top quark mass $M_t \lesssim 171$~GeV. In this case, it has been shown that the small values of $\yeff(\mu)$ near the Planck scale can significantly reduce the non-minimal coupling $\xi$ required for successful inflation~\cite{Bez09a,DeS09,Bez09b}.\footnote{Actually, the one-loop~\cite{Bez09a} and two-loop~\cite{DeS09,Bez09b} analyses predate the Higgs mass measurement and were carried out to determine the range of $M_h$ allowed for Higgs $\xi$-inflation. In retrospect, however, a Higgs mass near the lower end of the allowed region suggests a value of $\xi \lesssim 10^3$ is required for successful inflation, with the lower limit of $\xi$ unknown.} The reason for this is relatively simple: the tree-level estimate \eqref{eq:xi} shows that it is the combination $\lambda/\xi^2$ that must be small ($\sim 4 \times 10^{-10}$) in order to give the proper normalization of the CMB power spectrum. If $\yeff(\mu)$ is much smaller in the inflationary region than its tree-level value $\lambda \simeq 0.13$, then $\xi$ must also be smaller than the tree-level estimate $\xi \simeq 18000$.

The smaller value of $\xi$ required for successful inflation is particularly important since it is closely related to one of the most significant drawbacks of Higgs $\xi$-inflation: the violation of perturbative unitarity at the scale $\Mp/\xi$. For $\xi \rightarrow 1$, this scale is pushed toward the inflationary scale $\Mp/\sqrt{\xi}$ and the questionable assumptions of non-renormalizable operators~\cite{Ler10} or new strong dynamics~\cite{DeS09,Bez09b,Bez11,Bez11b} entering to restore unitarity are no longer required.\footnote{Of course, the Higgs field $h$ during inflation becomes trans-Planckian in this case and one must worry about the effects of higher dimensional operators suppressed by the Planck scale, which may spoil the flatness of the potential or the inflationary predictions~\cite{Bez09c}. As remarked in~\cite{DeS09}, however, the same worry applies to many minimal models of inflation, such as $m^2\phi^2$ chaotic inflation.} Since the lower limit of $\xi$ in the case of small $\yeff(\mu)$ during inflation has not been explored, an important question is whether it is possible to realize Higgs $\xi$-inflation with $\xi \sim 1$ and hence avoid the perturbative unitarity issues with the model. Such a region is, by nature, highly sensitive to the running of $\yeff(\mu)$ and requires a proper loop analysis within the Higgs $\xi$-inflation framework.

To investigate the lower limit of $\xi$ with $\yeff(\mu) \ll 1$ during inflation, we first describe the RG equations and the two-loop effective potential that are appropriate for Higgs $\xi$-inflation in sections~\ref{sec:RGE} and \ref{sec:potential}, respectively. The analysis of inflation, including the lower limits on $\xi$ and the inflationary predictions, are presented in section~\ref{sec:results}.

\subsection{Renormalization group equations}
\label{sec:RGE}
The modification of the well-known RG equations of the SM for the Higgs $\xi$-inflation scenario has been discussed in~\cite{Bez09a,DeS09,Bez09b,Cla09,Ler09,Ler11}. Essentially, the scalar propagator of the physical Higgs field, which enters into loop diagram calculations for the RG equations, must be multiplied by the field-dependent factor~\cite{DeS09,Ler09}\footnote{The reason for this suppression is that the canonical momentum of $h$ (which is evaluated in the Einstein frame with a canonical gravity sector) gives a non-standard commutator $[h(\vec x),\dot{h}(\vec y)] = i \hbar s(h)\delta^3(\vec x - \vec y)$ in the Jordan frame after imposing the standard commutation relations $[h(\vec x),\pi(\vec y)] = i\hbar \delta^3(\vec x - \vec y)$.}
\begin{equation}
\label{eq:s}
s(h) = \frac{1+\xi h^2/\Mp^2}{1+(1+6\xi)\xi h^2/\Mp^2}.
\end{equation}
For small field values $h \ll \Mp/\xi$, $s \simeq 1$ and the RG equations for the SM are perfectly adequate for describing the RG evolution. For large field values $h \gg \Mp/\xi$, however, the physical Higgs propagator is suppressed by a factor $s \simeq 1/(1+6\xi)$ and hence the RG equations differ from those of the SM. Two methods of dealing with this effect have been considered in the literature~\cite{DeS09,Bez09b}, leading to somewhat different results.

The first method of treating the suppressed Higgs loops, which is described in~\cite{DeS09}, is to insert one suppression factor $s$ into the RG equations of the SM for each off-shell Higgs propagator. Originally this was done by extracting out all Higgs doublet propagators at one-loop order and inserting the appropriate factors of $s$, repeating the process only for obvious terms at two-loop order~\cite{DeS09}. It was later pointed out, however, that only the propagator of the physical Higgs field and not the Nambu-Goldstone bosons that are present in the Landau gauge should come with such a factor~\cite{Bez09b}. The corrected RG equations with systematic insertions of $s$ for all two-loop terms, except for $\beta_\lambda$, are given in~\cite{Ler11}. By using these RG equations in the full two-loop SM effective potential from~\cite{Deg12} (with $m^2 \rightarrow 0$ and $M_h^2 \rightarrow 3s\lambda h^2$) and demanding that the potential be independent of $\mu$, we have been able to extract the two-loop part for $\beta_\lambda$.\footnote{In the process, we believe that two typos in the complete expression for the two-loop SM effective potential have been discovered. In~\cite{Deg12}, the final term of (A.3) should read $-\chi I_{ttg}$ instead of $-\chi I_{ttz}$ and the second last term on the third line of (A.5) should read $-3I_{w00}$ instead of $+I_{w00}$.} A similar procedure can then be used to obtain the two-loop RG equation for the Higgs mass parameter $m^2$ (in the notation of~\cite{Deg12}) with appropriate suppression factors. Although $\beta_{m^2}$ is not actually required for an analysis of Higgs $\xi$-inflation, it can be used to derive the RG equation for $\xi$ through the relation $\beta_\xi = \paren{\xi + 1/6}\gamma_m$, where $\gamma_m = \beta_{m^2}/m^2$~\cite{Ler09}. The complete set of two-loop RG equations with suppressed physical Higgs loops is given in appendix~\ref{app:RGE}.

The second method of treating the suppressed Higgs loops is to instead view the effect as a suppression of the effective Higgs coupling to other SM fields and, for large $\xi$, neglect the physical Higgs field  altogether in the region $h \gtrsim \Mp/\xi$~\cite{Bez09b}. The resulting theory (SM with frozen radial Higgs mode) is known as the chiral electroweak theory and has been studied previously in the literature. It is therefore possible to extract one-loop RG equations, which are valid for $\xi \gg 1$, from earlier works such as~\cite{Dut08}. In~\cite{Bez09b}, however, the RG equations for $\lambda$, $y_t$, and $\xi$ derived in this way differ from \eqref{eq:betalambda}, \eqref{eq:betayt}, and \eqref{eq:betaxi} with $s = 0$. A closer look reveals two sources (though not necessarily errors) for this discrepancy. First, the equation for the running of $v^2$ used in~\cite{Bez09b}, which was first given in~\cite{Dut08}, differs from the running of the SM Higgs field $h^2$ in the Landau gauge,
\begin{equation}
\label{eq:betah2}
16\pi^2\mu\frac{\partial}{\partial \mu}h^2 = \paren{\frac{3}{2}\gp{2}+\frac{9}{2}g^2-6y_t^2}h^2.
\end{equation}
In the usual SM case, the running of the Higgs vacuum expectation value $v^2$ and the Higgs field $h^2$ are both gauge-dependent quantities~\cite{Buc95} and have the same running~\cite{Ara92}. Although it has been argued that $v^2$ in the chiral electroweak theory is a gauge-invariant parameter and therefore its running should also be gauge invariant, we have found it difficult to reproduce the chiral electroweak theory result using Feynman diagrams and understand why its running differs from the running of $h^2$ in the SM. In any case, if eq.~(5.6) of~\cite{Bez09b} is replaced by the similar expression \eqref{eq:betah2}, the resulting one-loop equation for $\beta_{y_t}$ agrees with \eqref{eq:betayt} for $s = 0$. Second, $\beta_\lambda$ is derived in \cite{Bez09b} by demanding that the one-loop effective potential be independent of $\mu$, where the one-loop potential does not include the usual contribution from the Nambu-Goldstone bosons~\cite{Deg12}
\begin{equation}
\label{eq:deltaU1}
\Delta U_1 = \frac{3M_\GB^4}{64\pi^2}\paren{\ln\frac{M_\GB^2}{\mu^2}-\frac{3}{2}},
\end{equation}
where $M_\GB^2 = \lambda h^2$. Although the Goldstone boson contribution to the effective potential is strongly suppressed for prescription~I (see section~\ref{sec:potential}), the result of excluding this term in deriving $\beta_\lambda$ is equivalent to suppressing some Feynman diagrams with off-shell Goldstone boson propagators; that is, the $\paren{6+18s^2}\lambda^2$ term in \eqref{eq:betalambda} disappears entirely. While this difference is small in the region $h \gtrsim \Mp/\xi$ (numerically it is smaller than the two-loop correction to $\beta_\lambda$), if the contribution~\eqref{eq:deltaU1} is included in eq.~(4.1) of~\cite{Bez09b} the resulting one-loop equation for $\beta_\lambda$ agrees with \eqref{eq:betalambda} for $s = 0$. Note that the one-loop equation for $\beta_\xi$ in~\cite{Bez09b} still differs from \eqref{eq:betaxi} even after accounting for these changes. Specifically, the latter has a factor of $\xi + 1/6$ instead of $\xi$ and an additional term $\paren{6+6s}\lambda$ compared to the former. Again, these differences in $\beta_\xi$ are small (typically below the size of the two-loop correction to $\beta_\xi$) since we always have $\xi \gg 1/6$ and since $\lambda$ is small in the region $\Mp/\xi \lesssim h \lesssim \Mp/\sqrt\xi$ over which $\xi$ runs.

It is also worth mentioning that the second method includes the effects of additional counterterms taken from the chiral electroweak theory~\cite{Dut08} that arise to cancel divergences in the non-renormalizable SM sector without a Higgs field. These effects appear through the additional couplings $\alpha_0$ and $\alpha_1$ that modify the renormalization of the Z boson mass~\cite{Bez09b} and contribute to the effective potential at the two-loop level. Numerically, however, the Z boson mass contribution at the two-loop level, and hence this effect, is subleading~\cite{Deg12}.

The two methods of treating the suppressed Higgs loops for $h \gtrsim \Mp/\xi$ are therefore quite similar, at least for large $\xi$. The first method uses $s$ factors to smoothly interpolate between the SM-like RG evolution at low energies and the RG evolution with suppressed physical Higgs propagators at high energies, while the second method models this transition as an abrupt change at $\Mp/\xi$.\footnote{A smooth interpolation is preferred, but for $\xi \gg 1$ the function $s(h)$ changes rapidly in the region $h \sim \Mp/\xi$ and so the modification of the numerics is negligible~\cite{Bez09b}.} Since the $s$-factor treatment also handles the case $\xi \sim 1$, though, we adopt the first method~\cite{DeS09,Cla09,Ler09,Ler11} and use a suppression factor $s$ for each off-shell Higgs propagator in the SM RG equations for our analysis of Higgs $\xi$-inflation. Despite the different treatments of the RG equations described above, the numerical differences are small enough that a two-loop analysis in the region $h \gtrsim \Mp/\xi$ should be justified.

With the recent SM calculation of the three-loop beta functions for the gauge couplings~\cite{Mil12} and the leading three-loop terms for $\beta_\lambda$, $\beta_{y_t}$, and $\gamma$~\cite{Che12}, it is relatively simple to include these contributions in the RG equations \eqref{eq:betalambda}--\eqref{eq:gamma} so that the Higgs $\xi$-inflation analysis matches the NNLO analysis of~\cite{Deg12} for $h \lesssim \Mp/\xi$ (see appendix~\ref{app:RGE}). Note that we do not attempt to insert the appropriate factors of $s$ into these expressions since the corrections would be smaller than the uncertainty in the RG equations for $h \gtrsim \Mp/\xi$.

For a complete description of the RG evolution in Higgs $\xi$-inflation, the equations \eqref{eq:betalambda}--\eqref{eq:gamma} with the three-loop corrections \eqref{eq:deltabetalambda}--\eqref{eq:deltagamma} must be supplemented by values of the SM couplings at the electroweak scale and the value of $\xi$ at some high scale, say $\Mp/\xi$. Appropriate initial values for the SM couplings can be found in~\cite{Deg12}. For the central values of $\alpha_Y^{-1}(M_Z)$ and $\alpha_2^{-1}(M_Z)$, the initial values of the gauge couplings $g^\prime$ and $g$ are
\begin{align}
\label{eq:ICgp}
g^\prime(M_Z) &= \sqrt{\frac{4\pi}{98.35}} \simeq 0.3575, \\
\label{eq:ICg}
g(M_Z) &= \sqrt{\frac{4\pi}{29.587}} \simeq 0.65171,
\end{align}
where $M_Z = 91.1876$~GeV. For the strong gauge coupling $g_s$, the initial value depends more sensitively on the uncertainty in $\alpha_s(M_Z) = 0.1184 \pm 0.0007$. It is given by
\begin{equation}
\label{eq:ICgs}
g_s(M_t) = 1.1645 + 0.0031\paren{\frac{\alpha_s(M_Z)-0.1184}{0.0007}} -0.00046\paren{\frac{M_t}{\text{GeV}}-173.15},
\end{equation}
where $M_t$ is the top quark pole mass determined from experiment. The initial value of the top quark Yukawa coupling $y_t$ is
\begin{align}
\label{eq:ICyt}
y_t(M_t) &= 0.93587 + 0.00557\paren{\frac{M_t}{\text{GeV}}-173.15}-0.00003\paren{\frac{M_h}{\text{GeV}}-125}\nonumber \\ &\quad -0.00041\paren{\frac{\alpha_s(M_Z)-0.1184}{0.0007}} \pm 0.00200_\text{th},
\end{align}
where $M_h$ is the Higgs pole mass, while the initial value of the Higgs quartic coupling $\lambda$ is
\begin{equation}
\label{eq:ICy}
\lambda(M_t) = 0.12577 + 0.00205\paren{\frac{M_h}{\text{GeV}}-125} - 0.00004\paren{\frac{M_t}{\text{GeV}}-173.15} \pm 0.00140_\text{th}.
\end{equation}
Note that the theoretical uncertainty for $\lambda(M_t)$ in~\eqref{eq:ICy} is equivalent to an uncertainty in the Higgs pole mass of $\pm 0.7$~GeV~\cite{Deg12}. This means, in particular, that for the measured Higgs mass $M_h = 125.7 \pm 0.4$~\cite{Baa13} it is quite reasonable to use values of $M_h \simeq 124$--127~GeV in \eqref{eq:ICyt} and \eqref{eq:ICy}. For the non-minimal coupling $\xi$, we are (a priori) free to choose its initial value $\xi_0$ at some high scale. We take the scale to be $\Mp/\xi_0$ so that, by definition,
\begin{equation}
\label{eq:ICxi}
\xi(\Mp/\xi_0) = \xi_0.
\end{equation}

The RG equations \eqref{eq:betalambda}--\eqref{eq:deltagamma} with the initial values \eqref{eq:ICgp}--\eqref{eq:ICxi} are therefore the ones we use to describe the RG evolution of the couplings for Higgs $\xi$-inflation.

\subsection{Two-loop effective potential}
\label{sec:potential}
The effective potential for Higgs $\xi$-inflation, like the RG equations, differs from the well-known SM result~\cite{For92,Deg12} due to the suppression of the physical Higgs propagators. As described in~\cite{Bez09a}, however, the effective potential cannot be fixed unambiguously; there are two inequivalent renormalization prescriptions depending on whether quantum corrections to the potential are computed in the Einstein frame (prescription~I)~\cite{Bez08} or the Jordan frame (prescription~II)~\cite{Bar08}. Without knowing the behaviour of the quantum theory at the Planck scale, it is unclear which prescription should be used. The former prescription has been connected to ideas of a possible quantum scale invariance~\cite{Sha09a,Sha09b,Bez13} while~\cite{Bar08} has argued that the latter prescription is correct because the Jordan frame is the one in which physical distances are measured.

For sufficiently large $\yeff(\mu)$, the running of $\yeff(\mu)$ during inflation is small and the choice of renormalization prescription is irrelevant from a practical point of view~\cite{Bez09a,Bez09b,Deg12}. For the small $\yeff(\mu)$ allowed by the recent Higgs mass measurement, however, the choice of renormalization prescription can significantly affect the behaviour of $\yeff(\mu)$ and hence the potential during inflation. Both renormalization prescriptions must therefore be considered.

For prescription~I, the tree-level SM potential
\begin{equation}
\label{eq:Vtree}
V_0(h) = \frac{\lambda}{4}\paren{h^2-v^2}^2 \simeq \frac{\lambda}{4}h^2
\end{equation}
is first rewritten in the Einstein frame ($h = h(\chi)$) using \eqref{eq:potential}, giving
\begin{equation}
\label{eq:Utree}
U_0(\chi) = \frac{\lambda h^4}{ 4 \Omega^4}.
\end{equation}
Note that the $v^2$ term in \eqref{eq:Vtree} has been safely neglected in the inflationary region $h^2 \gtrsim \Mp^2/\xi \gg v^2$. The one-loop radiative corrections induced by the fields of the SM then take the Coleman-Weinberg form~\cite{Deg12}\footnote{Up to corrections from the time-dependence of the background Higgs field as it rolls down its potential. Such corrections have been considered in~\cite{Moo11,Geo12} for the simpler Abelian Higgs model but have not yet been studied for the Higgs $\xi$-inflation model. Analyzing these corrections goes beyond the scope of this paper.}
\begin{align}
U_1(\chi) &= \frac{1}{16\pi^2} \left[ \frac{3 M_W^4}{2} \paren{\ln \frac{M_W^2}{\mu^2} - \frac{5}{6} } + \frac{3 M_Z^4}{4} \paren{\ln \frac{M_Z^2}{\mu^2} - \frac{5}{6} } - 3 M_t^4 \paren{ \ln \frac{M_t^2}{\mu^2} - \frac{3}{2} } \right. \nonumber \\
&\quad + \left. \frac{M_h^4}{4} \paren{\ln \frac{M_h^2}{\mu^2} - \frac{3}{2} } + \frac{3M_\GB^4}{4} \paren{\ln \frac{M_\GB^2}{\mu^2} - \frac{3}{2} } \right], \label{eq:Uoneloop}
\end{align}
where the particle masses $M_W$, $M_Z$, $M_t$, $M_h$, and $M_\GB$ are computed from the tree-level potential~\eqref{eq:Utree}, giving~\cite{Bez08,Bez09a,Sal13}\footnote{We obtain this result by expanding $\Higgs = \frac{1}{\sqrt 2} \paren{\begin{smallmatrix} 0 \\ h \end{smallmatrix}} + \paren{\begin{smallmatrix} \hat{\GB}^+ \\ ( \hat{h} + i \hat{\GB}^0 ) / \sqrt 2 \end{smallmatrix}}$ in the full expression for the tree-level potential $U_0 = \lambda(\Higgs^\dag \Higgs)^2/\Omega^4 = \lambda(\Higgs^\dag \Higgs)^2/(1+2\xi \Higgs^\dag \Higgs / \Mp^2)^2$ to quadratic order in the fields $\hat{h}$, $\hat{\GB}^+$, $\hat{\GB}^0$, where $h$ is the classical background value of the Higgs field $\hat{h}$~\cite{Sal13}.}
\begin{alignat}{2}
M_W^2 &= \frac{g^2 h^2}{4 \Omega^2}, &\qquad M_Z^2 &= \frac{\paren{g^2 + \gp{2}}h^2}{4\Omega^2}, \qquad M_t^2 = \frac{y_t^2 h^2}{2 \Omega^2}, \nonumber \\
\label{eq:massesI} M_h^2 &= \frac{3s\lambda h^2}{\Omega^4}\parenfrac{1-\xi h^2/\Mp^2}{1+\xi h^2/\Mp^2}, &\qquad M_\GB^2 &= \frac{\lambda h^2}{\Omega^4}.
\end{alignat}
Note that the particle masses $M_W^2$, $M_Z^2$, and $M_t^2$ in \eqref{eq:massesI} differ from the flat space results by the conformal factor $\Omega^2 = 1 + \xi h^2/\Mp^2$ that appears in the denominator, while the physical Higgs mass $M_h^2$ and Goldstone boson mass $M_\GB^2$ contain additional factors. With the exception of the suppression factor $s = s(h)$ in the physical Higgs mass, the appearance of these additional factors in $M_h^2$ and $M_\GB^2$ is due to using the asymptotically flat tree-level potential~\eqref{eq:Utree} to determine particle masses rather than the Jordan frame potential~\eqref{eq:Vtree}. These additional factors lead to a suppression of the physical Higgs and Goldstone boson contributions to the effective potential (relative to those from $W$, $Z$, and $t$) during inflation for prescription~I, as found in~\cite{Bez08,Bez09a,Sal13}.

The two-loop radiative corrections $U_2(\chi)$ can easily be found by using the modified particle masses \eqref{eq:massesI} in the two-loop SM result of~\cite{Deg12}, but due to the rather long and unenlightening form of this expression we do not reproduce it here. The RG-improved effective potential is then determined from $U_\text{eff}(\chi) = U_0 + U_1 + U_2$ in the usual way by using the RG equations from appendix~\ref{app:RGE} to run the couplings and making the replacement $h \rightarrow e^{\Gamma(\mu)} h$, where
\begin{equation}
\label{eq:anomalousdimension}
\Gamma(\mu) = -\int_{M_t}^{\mu} \gamma(\mu^\prime) d \ln \mu^\prime
\end{equation}
and $\gamma = -d\ln h/d\ln\mu$ is the anomalous dimension of the Higgs field~\cite{For93}.\footnote{Note the difference in sign between the definition of $\gamma$ here and the definition of $\gamma$ in~\cite{Deg12}.} The effective Higgs quartic coupling $\yeff(\mu)$ is then defined through
\begin{equation}
\label{eq:Ueff}
U_\text{eff}(\chi) \equiv \frac{\yeff(\mu)h^4}{4\Omega^4},
\end{equation}
where all couplings in \eqref{eq:Ueff} are evaluated at some renormalization scale $\mu$. The dependence of the effective potential on the scale $\mu$ is spurious, but to minimize the logarithms from higher loop corrections it is appropriate to take $\mu = \kappa h / \Omega$ proportional to the background mass of a vector boson or top quark~\cite{Bez09b}. For simplicity, we choose the constant of proportionality to be $\kappa = 1$.

For prescription~II, quantum corrections to the potential~\eqref{eq:Vtree} are computed in the Jordan frame before transforming to the Einstein frame. In this case, the one-loop radiative corrections to the effective potential take the form~\cite{Ler09}
\begin{align}
U_1(\chi) &= \frac{1}{16\pi^2\Omega^4} \left[ \frac{3 M_W^4}{2} \paren{\ln \frac{M_W^2}{\mu^2} - \frac{5}{6} } + \frac{3 M_Z^4}{4} \paren{\ln \frac{M_Z^2}{\mu^2} - \frac{5}{6} } - 3 M_t^4 \paren{ \ln \frac{M_t^2}{\mu^2} - \frac{3}{2} } \right. \nonumber \\
&\quad + \left. \frac{M_h^4}{4} \paren{\ln \frac{M_h^2}{\mu^2} - \frac{3}{2} } + \frac{3M_\GB^4}{4} \paren{\ln \frac{M_\GB^2}{\mu^2} - \frac{3}{2} } \right],
\end{align}
where the particle masses $M_W^2$, $M_Z^2$, $M_t^2$, $M_h^2$, and $M_\GB^2$ appear without the conformal factor $\Omega^2$ or additional factors in their denominators,
\begin{alignat}{2}
M_W^2 &= \frac{g^2 h^2}{4}, &\qquad M_Z^2 &= \frac{\paren{g^2 + \gp{2}}h^2}{4}, \qquad M_t^2 = \frac{y_t^2 h^2}{2}, \nonumber \\
\label{eq:massesII} M_h^2 &= 3s\lambda h^2, &\qquad M_\GB^2 &= \lambda h^2.
\end{alignat}
The two-loop radiative corrections $U_2(\chi)$ can be found by using the particle masses~\eqref{eq:massesII} in the two-loop SM result~\cite{Deg12} and dividing the expression by the conformal factor $\Omega^4$. The effective Higgs quartic coupling $\yeff(\mu)$ is again defined through~\eqref{eq:Ueff}, but in this case taking the renormalization scale to be proportional to the background mass of a vector boson or top quark requires $\mu = \kappa h$. For simplicity, we again choose $\kappa = 1$.

In practice, the most significant difference between the effective potentials for the two renormalization prescriptions is the functional dependence $\mu = \mu(h)$.\footnote{The additional suppression of the physical Higgs and Goldstone boson masses for prescription~I is relatively minor since these masses, and hence their contributions to the effective potential, are small compared to $M_W$, $M_Z$, and $M_t$ for the small $\lambda$ in the inflationary region.} For prescription~I, $\mu = h / \Omega$ approaches a constant value in the inflationary region $h \gtrsim \Mp/\sqrt \xi$ (and hence so do the couplings $g(\mu)$, $g^\prime(\mu)$, etc.\ in \eqref{eq:massesI}) while for prescription~II the renormalization scale $\mu = h$ does not. As a result, the effective potential for prescription~I approaches a constant value in the inflationary region (even after including radiative corrections) while the effective potential for prescription~II, due to the continued running of the couplings, does not. This difference, as we will see, can have a large impact on Higgs $\xi$-inflation and its predictions for small $\yeff(\mu)$.

\section{Numerical results}
\label{sec:results}
\begin{figure}
\begin{center}
\includegraphics[scale=1]{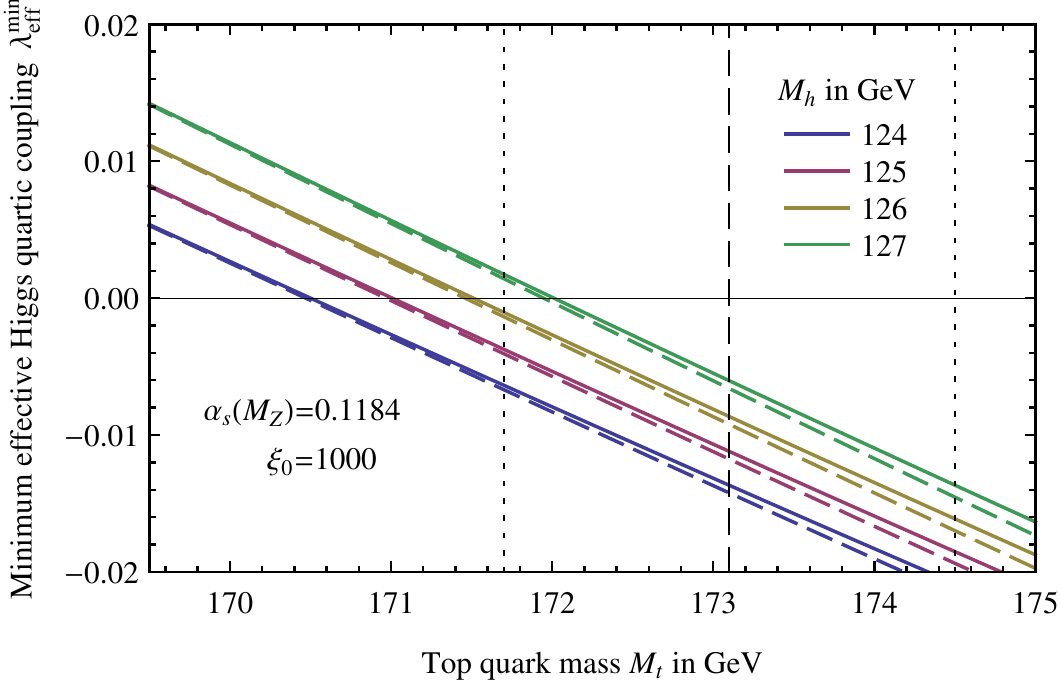}
\end{center}
\caption{Values of $\ymin$ as a function of $M_t$ for fixed $\alpha_s(M_Z) = 0.1184$ and $\xi_0 = 1000$. The four solid (dashed) curves correspond to a Higgs mass $M_h$ of 124, 125, 126, and 127~GeV from bottom to top for renormalization prescription~I~(II). The vertical dashed and dotted lines give the central value and $\pm 2\sigma$ range for $M_t$~\cite{Deg12}. A shift in $\alpha_s(M_Z)$ of $\pm 1\sigma$ ($\pm 0.0007$) roughly corresponds to a shift in $M_h$ of $\pm 0.5$~GeV while changing $\xi_0$ by an order of magnitude has little effect.}
\label{fig:Mt}
\end{figure}

For a fixed Higgs mass $M_h$, top quark mass $M_t$, strong coupling $\alpha_s(M_Z)$, and non-minimal coupling $\xi_0$, it is straightforward to numerically solve the RG equations \eqref{eq:betalambda}--\eqref{eq:deltagamma} with initial conditions \eqref{eq:ICgp}--\eqref{eq:ICxi} and use the effective potential $U(\chi)$ (for either prescription~I or II) to compute the inflationary parameters. However, since the focus of this paper is on the region of parameter space with $\yeff(\mu) \ll 1$, we instead replace the parameter $M_t$ in favour of $\ymin \equiv \min \{ \yeff(\mu) \}$. Intuitively, this can be understood as adjusting the top quark mass $M_t$ to yield the desired $\ymin$ for a fixed choice of $M_h$, $\alpha_s(M_Z)$, and $\xi_0$. Figure~\ref{fig:Mt} shows that the special region $\ymin \simeq 0$ exists for a top quark mass $M_t \sim 171$~GeV about 2--3$\sigma$ below its central value. Since values of $\ymin \sim 0.01$ are typical within the experimental and theoretical uncertainty of the various parameters, a fine-tuning of some combination of parameters is necessary to achieve $0 < \ymin \lesssim 0.01$.\footnote{In~\cite{Sha10} it is argued that a UV fixed point in an asymptotically safe theory of gravity may ensure very small values of $\lambda(\mu)$ near the Planck scale. In this case, fine-tuning may only be necessary for values of $\ymin$ smaller than the typical size of the shift in $\yeff(\mu)$ due to radiative corrections to the effective potential, $\delta \yeff(\mu \sim \Mp) \sim 4 \times 10^{-4}$.} Note that negative values of $\ymin$, as well as sufficiently small positive values, cause the effective potential to develop a second minimum below the inflationary scale and hence spoil Higgs $\xi$-inflation. We therefore restrict ourselves to the region $0 < \ymin \lesssim 0.01$ in which the effective potential is stable.

The non-minimal coupling $\xi_0$ is not actually a free parameter, of course, but must be chosen to give the correct normalization of the CMB power spectrum (see section~\ref{sec:review}).
For a fixed $M_h$ and $\alpha_s(M_Z)$, the procedure for determining the inflationary predictions for a particular choice of $\ymin$ and renormalization prescription is as follows:
\begin{enumerate}
\item Choose a value of $\xi_0$. Adjust the top quark mass $M_t$ to give the desired value of $\ymin$ when solving the RG equations. For $\ymin \lesssim 0.01$, this may involve fine-tuning $M_t$.
\item Use the effective potential $U(\chi)$ (for prescription~I or II) to compute the inflationary parameters and determine $U(h_0)/\epsilon(h_0)$ at a field value $h_0$ corresponding to $N = 59$ e-folds before the end of inflation.
\item Repeat the steps above for different values of $\xi_0$ until the correct normalization $U/\epsilon \simeq \paren{0.0274\Mp}^4$ is achieved.\footnote{Note that prescription~II with sufficiently small $\ymin$ can have two solutions for $\xi_0$. The large $\xi_0$ solution, however, predicts $n_s > 1.02$ and is therefore inconsistent with observations~\cite{WMA12,Pla13}.}
\item Compute the inflationary predictions for the spectral index $n_s$, the tensor-to-scalar ratio $r$, and the running of the spectral index $dn_s/d\ln k$.
\end{enumerate}
We discuss the numerical results for prescriptions~I and II separately.

\subsection{Inflationary predictions for prescription I}
\begin{figure}
\begin{center}
\includegraphics[scale=1]{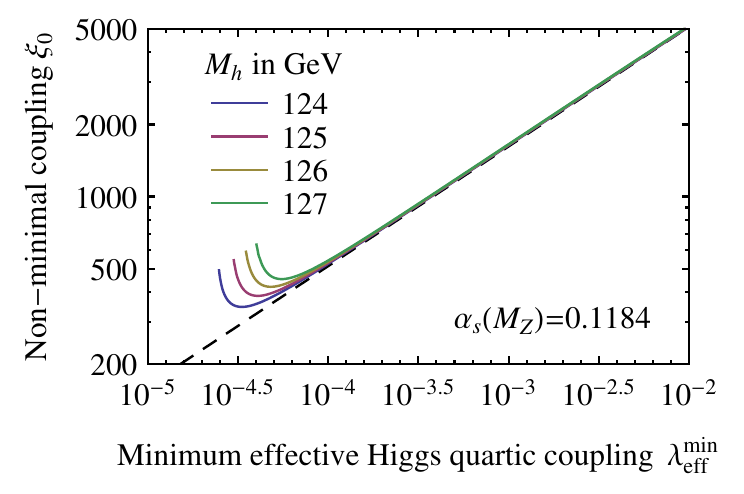}
\includegraphics[scale=1]{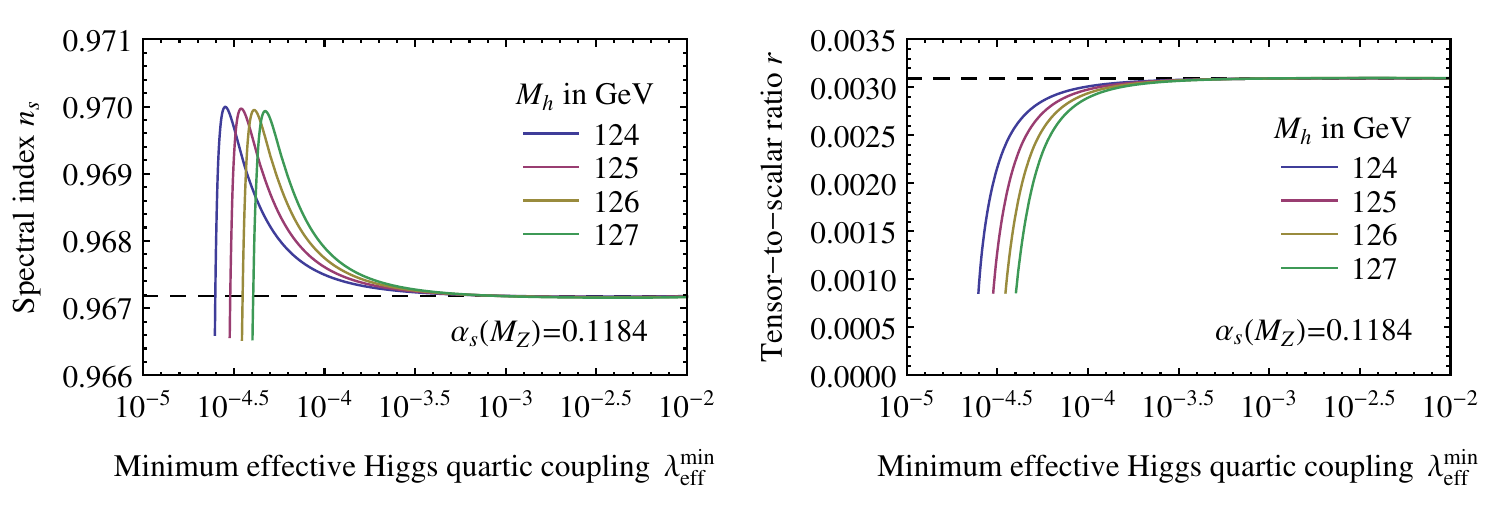}
\end{center}
\caption{Numerical results for the non-minimal coupling $\xi_0$ and inflationary predictions for the spectral index $n_s$ and the tensor-to-scalar ratio $r$ as a function of $\ymin$ for prescription~I. The four solid curves correspond to a Higgs mass $M_h$ of 124, 125, 126, and 127~GeV from left to right while the dashed lines give the tree-level predictions. A shift in $\alpha_s(M_Z)$ of $\pm 2\sigma$ ($\pm 0.0014$) roughly corresponds to a shift in $M_h$ of $\mp 0.5$~GeV. Changing the number of e-folds from $N = 59$ to 62 shifts the tree-level predictions by a small (calculable) amount but does not change the qualitative behaviour of the curves about the tree-level predictions.}
\label{fig:resultsI}
\end{figure}

For prescription~I, the results for $\xi_0$ and the inflationary predictions for $n_s$ and $r$ (as a function of $\ymin$) are presented in figure~\ref{fig:resultsI}. The running of the spectral index $dn_s/d\ln k$ always remains small, within the range $(5.0\text{--}5.6) \times 10^{-4}$.

Let us first discuss the non-minimal coupling $\xi_0$. Figure~\ref{fig:resultsI} shows that the value of $\xi_0$ required for the CMB normalization deviates from the tree-level estimate $\xi_0 \simeq 48000\sqrt{\ymin}$ as $\ymin$ decreases below about $10^{-4}$. In particular, $\xi_0$ reaches a minimum value of $\xi_0 \sim 400$ at $\ymin \sim 10^{-4.4}$ and then begins to increase. This behaviour can be traced to the rapid decrease in $\epsilon$ (and hence the tensor-to-scalar ratio $r = 16\epsilon$) over this range, which causes $U/\epsilon$ to increase despite smaller values of $\ymin$. A larger non-minimal coupling $\xi_0$ is therefore required to give the correct CMB normalization. This result demonstrates that the sharp decrease in $\xi_0$ seen in~\cite{Bez09a} and figure~4 of~\cite{Bez09b} for prescription~I does not continue indefinitely but only allows $\xi_0$ as small as about 400. The violation of perturbative unitarity at the scale $\Mp/\xi_0 \ll \Mp / \sqrt{\xi_0}$ therefore remains a problem for Higgs $\xi$-inflation in the small $\ymin$ region. For sufficiently small $\ymin$ (e.g.\ $\lesssim 10^{-4.6}$), no solutions for $\xi_0$ are possible since the effective potential develops a second minimum and hence spoils the Higgs $\xi$-inflation scenario.\footnote{In this case, the effective potential rises to a local maximum and then decreases slowly to a constant value as $h \rightarrow \infty$. The shape of the potential may be suitable for a sort of false vacuum inflation in which the Higgs field can start with any value $h \gtrsim \Mp/\sqrt{\xi}$, but an analysis of this case goes beyond the scope of this paper.}

Figure~\ref{fig:resultsI} also shows small deviations in the inflationary predictions for the spectral index $n_s$ and the tensor-to-scalar ratio $r$. As $\ymin$ decreases below about $10^{-3.5}$, the spectral index rises to about 0.970 from its tree-level prediction of 0.967 before decreasing rapidly, while the tensor-to-scalar ratio drops quickly below its tree-level prediction of 0.0031. Although a similarly rapid change in $n_s$ and $r$ can be seen in~\cite{Bez09a,Bez09b} as the Higgs mass approaches values corresponding to $\ymin \simeq 0$, the results presented here (as a function of $\ymin$) provide a much clearer picture of Higgs $\xi$-inflation in this now experimentally favoured region. From a practical point of view, we see that the deviations of $n_s$ and $r$ from the tree-level predictions are sufficiently small that they would be difficult to distinguish from the tree-level results observationally. Consequently, for all allowed values $10^{-4.6} \lesssim \ymin \lesssim 10^{-2}$, the predictions for $n_s$ and $r$  are well within the current 1$\sigma$ limits~\cite{WMA12,Pla13}. The small prediction for $dn_s/d\ln k \sim 5 \times 10^{-4}$ is also consistent with observations at the 1--2$\sigma$ level~\cite{WMA12,Pla13}.

\subsection{Inflationary predictions for prescription II}
\begin{figure}
\begin{center}
\includegraphics[scale=1]{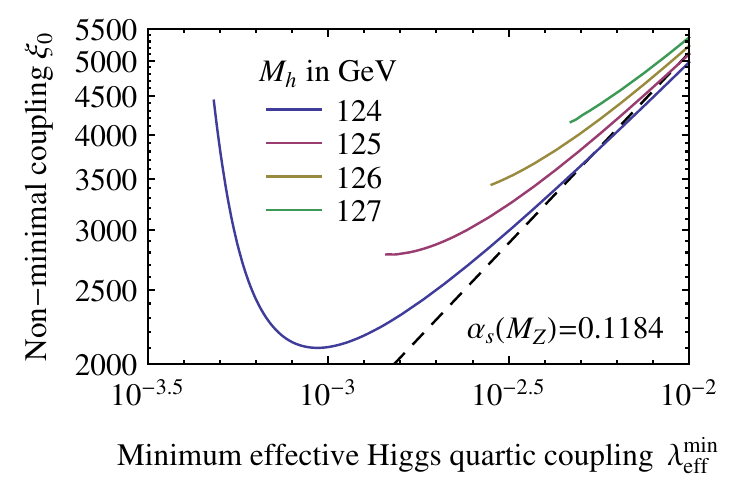}
\includegraphics[scale=1]{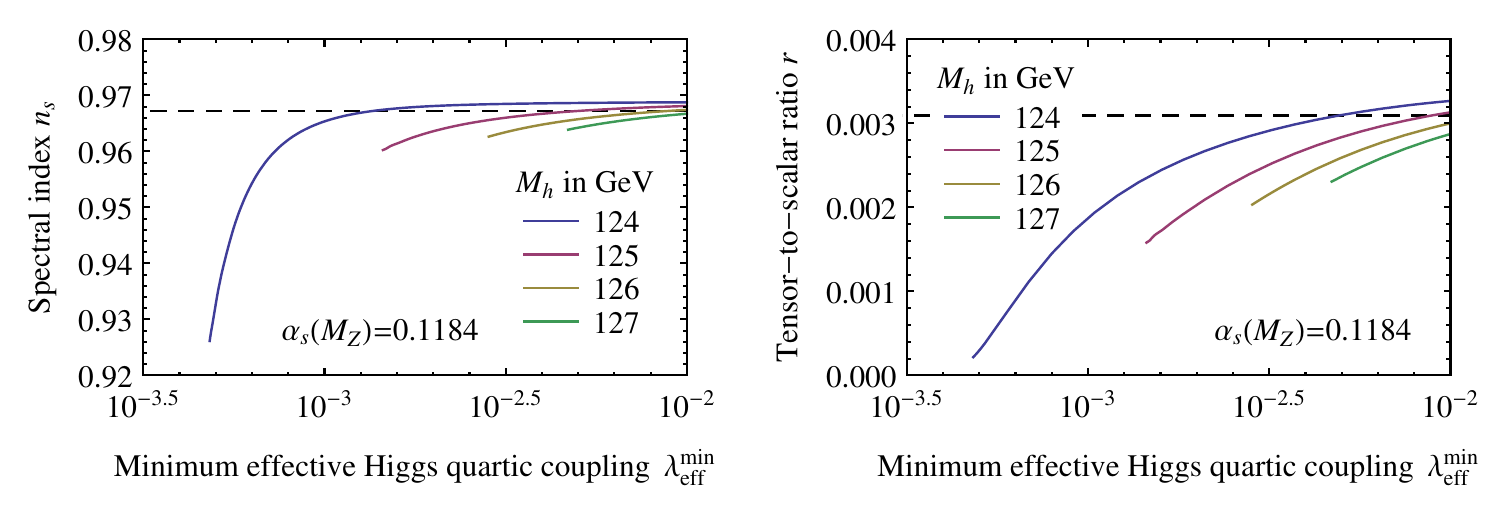}
\end{center}
\caption{Numerical results for the non-minimal coupling $\xi_0$ and inflationary predictions for the spectral index $n_s$ and the tensor-to-scalar ratio $r$ in the larger $\ymin$ region for prescription~II. The four solid curves correspond to a Higgs mass $M_h$ of 124, 125, 126, and 127~GeV from left to right while the dashed lines give the tree-level predictions. A shift in $\alpha_s(M_Z)$ of $\pm 2\sigma$ ($\pm 0.0014$) roughly corresponds to a shift in $M_h$ of $\mp 0.5$~GeV. Changing the number of e-folds from $N = 59$ to 62 shifts the tree-level predictions by a small (calculable) amount but does not change the qualitative behaviour of the curves about the tree-level predictions.}
\label{fig:resultsII}
\end{figure}

\begin{figure}
\begin{center}
\includegraphics[scale=1]{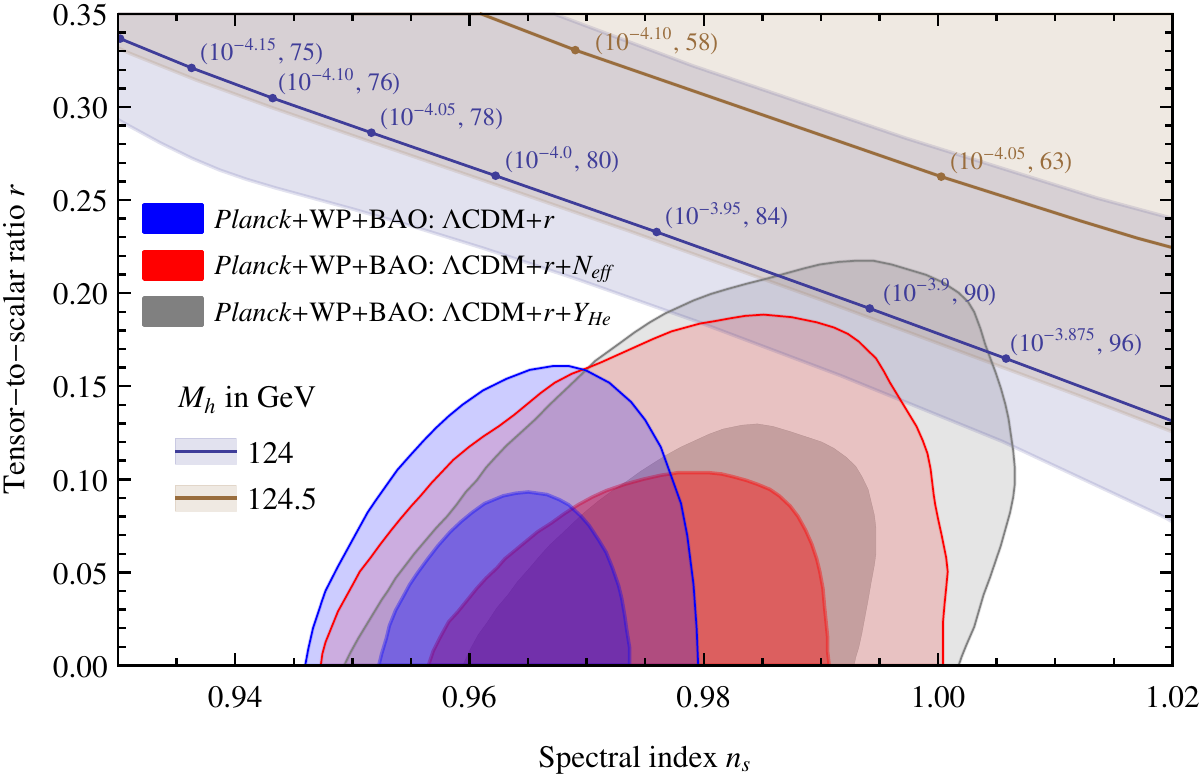}
\end{center}
\caption{Predictions for the spectral index $n_s$ and the tensor-to-scalar ratio $r$ for Higgs $\xi$-inflation with prescription~II and $\ymin \sim 10^{-4}$. The solid blue (brown) curve gives the results for a Higgs mass $M_h = 124$~GeV (124.5~GeV) while the lower and upper shaded regions correspond to a shift in $\alpha_s(M_Z) = 0.1184$ of up to $\pm 2\sigma$ ($\pm 0.0014$), respectively. The marked points along the solid curves indicate values of $(\ymin,\xi_0)$. Results are shown with the marginalized joint 68\% and 95\% confidence level regions from {\it Planck} 2013~\cite{Pla13}.}
\label{fig:lowII}
\end{figure}

For prescription~II, there are two disjoint regions of $\ymin$ that can lead to acceptable inflation: one with larger values $\ymin \gtrsim 10^{-3.3}\text{--}10^{-2.3}$ (depending on $M_h$) and one with smaller values $\ymin \sim 10^{-4}$. The results for $\xi_0$ and the inflationary predictions for $n_s$ and $r$ are quite different for these two regions and are presented in figures~\ref{fig:resultsII} and \ref{fig:lowII}, respectively.

Let us first consider the region of larger values of $\ymin$, which is the only one that has been considered previously in the literature~\cite{DeS09,Bez09a,Bez09b}. Figure~\ref{fig:resultsII} shows that the required value of $\xi_0$ in this region behaves similarly to that of prescription~I except that the minimum value of $\xi_0$ --- if it can be reached without the potential developing a second minimum --- occurs at larger $\ymin$ (i.e.\ $\ymin \gtrsim 10^{-3}$). This difference is due to the stronger effect of the running of $\yeff(\mu)$  for prescription~II. Specifically, the running of $\yeff(\mu)$ to its minimum value overcomes the flattening of the potential in the inflationary region more quickly than for prescription~I, and hence causes the effective potential to develop a second minimum for more moderate values of $\ymin$. As a result, a non-minimal coupling only as small as $\xi_0 \sim 2000$--4000 (depending on $M_h$) is allowed for this region of $\ymin$. Similar lower limits for $\xi_0$, as well as the qualitative rise in $\xi_0$ as $\ymin$ decreases, have been found in~\cite{Bez09a,Bez09b} for prescription~II. Again, these values of $\xi_0$ are not small enough to prevent the perturbative unitarity violation at $\Mp/\xi_0$ from occurring well below the inflationary scale.

Figure~\ref{fig:resultsII} also shows that the predictions for the spectral index $n_s$ and the tensor-to-scalar ratio $r$ \emph{decrease} from their tree-level values as $\ymin \rightarrow 0$. The decrease observed here (similar to prescription~I) is consistent with the results of~\cite{Bez09a,Bez09b} rather than with the increase observed in~\cite{DeS09}. Also note that the variation in $n_s$ over the allowed range of $\ymin$ is larger for prescription~II than for prescription~I. Since a deviation from the tree-level prediction of $\Delta n_s \gtrsim 0.01$ should be visible by {\it Planck}~\cite{Bur10b}, it may therefore be possible to connect a measurement of the spectral index with the RG evolution of $\yeff(\mu)$ near the Planck scale for prescription~II. The running of the spectral index $dn_s/d\ln k$ always remains quite small, within the range $(4.5\text{--}6.4) \times 10^{-4}$.

While the results of the larger $\ymin$ region for prescription~II are qualitatively similar to those for prescription~I, prescription~II also allows a region of smaller $\ymin$ and $\xi_0$ with distinct inflationary predictions. The existence of this region, which has not been considered in the literature before, can be understood as follows. For typical Higgs $\xi$-inflation with large $\ymin$, the slow roll parameter $\epsilon$ decreases rapidly in the inflationary region (see eq.~\eqref{eq:epsilon}) and the required $N = 59$ e-folds of inflation are produced quickly (see eq.~\eqref{eq:N}). For smaller $\ymin$ and $\xi_0$, however, there is a region of parameter space in which the running of $\yeff(\mu)$ causes $\epsilon$ to increase before the $N=59$ e-folds are reached. The inflationary observables are then computed at a field value $h_0$ in a qualitatively different region of parameter space with larger $\epsilon$, leading to distinct predictions.

Figure~\ref{fig:lowII} gives the numerical results for $\xi_0$ and the inflationary predictions for the spectral index $n_s$ and the tensor-to-scalar ratio $r$ in this region. The results are shown together with the most recent constraints from {\it Planck}~\cite{Pla13}. Since many well-motivated models with Higgs $\xi$-inflation contain additional degrees of freedom that can contribute to the effective number of neutrino species $N_\text{eff}$ (e.g.\ the $\nu$MSM~\cite{Asa05,Sha05} with 3 light sterile neutrinos), it is most appropriate to compare the results with the $\Lambda$CDM+$r$+$N_\text{eff}$ data. It can be seen that, for $\ymin \sim 10^{-3.9}$ and $M_h \simeq 124$~GeV, there is a region of Higgs $\xi$-inflation that is consistent with {\it Planck} at the 2--3$\sigma$ level.\footnote{Recall that even with the Higgs mass measurement of $M_h = 125.7 \pm 0.4$~\cite{Baa13}, using $M_h \simeq 124$~GeV in the RG evolution is still quite reasonable due to the theoretical uncertainty in determining $\lambda$ at the electroweak scale.}
This region, though marginally disfavoured, is important for two reasons. First, unlike for the larger $\ymin$ region, the tensor-to-scalar ratio $r \gtrsim 0.15$ in this region is quite large and would be visible by {\it Planck}~\cite{Bur10b}. It is therefore possible that the tensor modes from Higgs $\xi$-inflation could be detected in the {\it Planck} polarization data.\footnote{The running of the spectral index in this region is also much larger (and negative) than typically found in Higgs $\xi$-inflation: $dn_s/d\ln k \sim -0.002$ and $-0.008$ for $M_h = 124$ and 124.5~GeV, respectively. A large running of the spectral index relaxes the constraints on $r$ from {\it Planck}~\cite{Pla13} and could open up even more of the small $\ymin$ region for detection.}
Second, the non-minimal coupling $\xi_0 \sim 90$ required in this region is about an order of magnitude smaller than previously considered in the literature. Although still not small enough to address the problem of perturbative unitarity violation occurring below the inflationary scale, it provides the lower limit on $\xi_0$ that is acceptable for Higgs $\xi$-inflation. Smaller non-minimal couplings, including $\xi_0 \sim 1$, seem generally unattainable because they require $\ymin \lesssim 10^{-6}$ to give the correct CMB normalization, which ultimately causes the effective potential to develop a second minimum before the inflationary scale.

\section{Conclusions}
\label{sec:conclusions}
Higgs $\xi$-inflation is an attractive model of inflation since it does not require scalar degrees of freedom in addition to those of the SM. For a large non-minimal coupling $\xi$, however, the violation of perturbative unitarity at the scale $\Mp/\xi \ll \Mp$ threatens the self-consistency of the model in the inflationary region. In this paper we have investigated the possibility that a Higgs mass $M_h \simeq 125$--126~GeV --- a mass for which the effective Higgs quartic coupling $\yeff(\mu)$ runs to very small values near the Planck scale --- may significantly reduce the size of $\xi$ required for inflation and address the perturbative unitarity violation problem. This possibility, like the Higgs $\xi$-inflation scenario in general, requires a top quark mass $M_t \sim 171$~GeV, about 2$\sigma$ below its central value.

To investigate this possibility we have updated the two-loop analysis of Higgs $\xi$-inflation to include the three-loop SM beta functions for the gauge couplings as well as the leading three-loop terms for the RG evolution of $\lambda$, the top Yukawa coupling $y_t$, and the Higgs anomalous dimension $\gamma$. We have also included, for the first time, a complete two-loop insertion of suppression factors for the physical Higgs loops in the RG equations. The two-loop SM effective potential with particle masses modified appropriately for Higgs $\xi$-inflation has been used to match the level of the RG equations.

We have found that successful inflation in the region $\yeff(\mu) \ll 1$ requires smaller $\xi$ than previously considered in the literature, but even with a fine-tuning of parameters to give arbitrarily small $\ymin$ it is not possible to achieve $\xi \sim 1$ and prevent the violation of perturbative unitarity below the inflationary scale. Specifically, we have found that the Einstein frame renormalization prescription (prescription~I) allows a non-minimal coupling as small as $\xi \sim 400$ for $\ymin \sim 10^{-4.4}$ without the potential developing a second minimum and hence spoiling inflation. The predictions for the spectral index $n_s$ and the tensor-to-scalar ratio $r$ remain close to their tree-level values in this case and are within the 1$\sigma$ allowed region from CMB measurements. For the Jordan frame renormalization prescription (prescription~II), there are two distinct regions of $\ymin$ that can lead to successful inflation. The larger $\ymin$ region behaves similarly to prescription~I and allows a non-minimal coupling as small as $\xi \sim 2000$ without the potential developing a second minimum. The smaller $\ymin$ region, which has not been considered in the literature before, requires $\xi \sim 90$ and predicts an observable tensor-to-scalar ratio $r \gtrsim 0.15$ for $\ymin \sim 10^{-3.9}$. Smaller non-minimal couplings, including $\xi \sim 1$, seem generally unattainable since they require $\ymin \lesssim 10^{-6}$ to give the correct CMB normalization, which ultimately causes the effective potential to develop a second minimum before the inflationary scale.

\acknowledgments
I am grateful to Graham Ross for much valuable input and to Subir Sarkar for helpful discussions. I would also like to thank Andrei Linde, Mikhail Shaposhnikov, and Alberto Salvio for valuable correspondence. This work was supported by the European Commission under the Marie Curie Initial Training Network UNILHC 237920 (Unification in the LHC era). Contents reflect only the author's views and not the views of the European Commission.

\appendix
\section{Renormalization group equations for Higgs $\xi$-inflation}
\label{app:RGE}
In this appendix we list the (gauge-independent) RG equations for the couplings $\lambda$, $y_t$, $g^\prime$, $g$, $g_s$, and $\xi$ in the $\overline{\text{MS}}$ scheme that are used in our analysis of Higgs $\xi$-inflation. For each coupling we write $dx/dt = \beta_x$, where $t = \ln(\mu/\mu_0)$. The anomalous dimension of the Higgs field $\gamma$ in the Landau gauge, for use in \eqref{eq:anomalousdimension}, is also given. As described in section~\ref{sec:RGE}, the RG equations contain one suppression factor $s = s(h)$ for each off-shell physical Higgs propagator.\footnote{The suppression factor $s(h) = s(h(\mu))$ can be written in terms of $\mu$ by inverting $\mu = h / \Omega$ or $\mu = h$ for prescriptions~I or II, respectively (see section~\ref{sec:potential}).}
Note that the RG equations for the SM can be recovered by setting $s = 1$.

The two-loop RG equations for $\lambda$, $y_t$, $g^\prime$, $g$, $g_s$, and $\xi$ are as follows. For the Higgs quartic coupling we have
\begin{align}
\label{eq:betalambda}
\beta_\lambda &= \frac{1}{\paren{4\pi}^2}\left[ \paren{6 + 18 s^2}\lambda^2 - 6 y_t^4 + \frac{3}{8}\paren{2 g^4 + \paren{g^2 + \gp{2}}^2} + \paren{-9 g^2 - 3\gp{2} + 12 y_t^2}\lambda \right]\nonumber\\
&+ \frac{1}{\paren{4\pi}^4}\left[\frac{1}{48}\paren{\paren{912+3s}g^6-\paren{290-s}g^4 \gp{2} - \paren{560-s}g^2 \gp{4} - \paren{380-s}\gp{6}} \right. \nonumber\\
&+\paren{38-8s}y_t^6 - y_t^4\paren{\frac{8}{3}\gp{2} + 32 g_s^2 + \paren{12-117s+108s^2}\lambda}\nonumber\\
&+\lambda \left( -\frac{1}{8}\paren{181+54s-162s^2} g^4 + \frac{1}{4}\paren{3-18s+54s^2}g^2 \gp{2} + \frac{1}{24}\paren{90+377s+162s^2}\gp{4} \right. \nonumber\\
&+ \left. \paren{27+54s+27s^2}g^2 \lambda + \paren{9+18s+9s^2}\gp{2}\lambda-\paren{48+288s-324s^2+624s^3-324s^4}\lambda^2 \frac{}{} \right)\nonumber\\
&+ \left. y_t^2\paren{-\frac{9}{4}g^4 + \frac{21}{2}g^2 \gp{2} -\frac{19}{4}\gp{4}+\lambda\paren{\frac{45}{2}g^2+\frac{85}{6}\gp{2}+80g_s^2-\paren{36+108s^2}\lambda}} \right].
\end{align}
For the top quark Yukawa coupling we have
\begin{align}
\label{eq:betayt}
\beta_{y_t} &= \frac{y_t}{\paren{4\pi}^2}\left[-\frac{9}{4}g^2-\frac{17}{12}\gp{2}-8g_s^2+\paren{\frac{23}{6}+\frac{2}{3}s}y_t^2 \right] \nonumber\\
&+ \frac{y_t}{\paren{4\pi}^4}\left[ -\frac{23}{4}g^4 - \frac{3}{4}g^2 \gp{2} + \frac{1187}{216}\gp{4} + 9g^2 g_s^2 + \frac{19}{9}\gp{2}g_s^2 - 108 g_s^4  \right. \nonumber\\
&+ \left. \paren{\frac{225}{16}g^2 + \frac{131}{16}\gp{2} + 36 g_s^2}s y_t^2 + 6 \paren{-2s^2 y_t^4 - 2s^3 y_t^2 \lambda + s^2 \lambda^2} \right].
\end{align}
For the gauge couplings $g^\prime$, $g$, and $g_s$ we have
\begin{align}
\beta_{g^\prime} &= \frac{\gp{3}}{\paren{4\pi}^2} \squareparen{\frac{81+s}{12}} + \frac{\gp{3}}{\paren{4\pi}^4}\left[ \frac{199}{18} \gp{2} + \frac{9}{2}g^2 + \frac{44}{3}g_s^2 - \frac{17}{6}s y_t^2 \right],\label{eq:betagp}
\end{align}
\begin{align}
\beta_g &= \frac{g^3}{\paren{4\pi}^2} \squareparen{-\frac{39-s}{12}} + \frac{g^3}{\paren{4\pi}^4}\left[ \frac{3}{2} \gp{2} + \frac{35}{6}g^2 + 12 g_s^2 - \frac{3}{2}s y_t^2 \right],\label{eq:betag}
\end{align}
\begin{align}
\beta_{g_s} &= \frac{g_s^3}{\paren{4\pi}^2} \squareparen{-7} + \frac{g_s^3}{\paren{4\pi}^4}\left[ \frac{11}{6} \gp{2} + \frac{9}{2}g^2 -26 g_s^2 - 2s y_t^2 \right].\label{eq:betags}
\end{align}
And for the non-minimal coupling $\xi$ we have
\begin{align}
\label{eq:betaxi}
\beta_\xi &= \frac{1}{\paren{4\pi}^2} \paren{\xi + \frac{1}{6}} \left[ -\frac{3}{2}\gp{2}-\frac{9}{2}g^2+6y_t^2 + \paren{6+6s}\lambda\right] \nonumber\\
&+\frac{1}{\paren{4\pi}^4} \paren{\xi + \frac{1}{6}}\left[\paren{-\frac{199}{16}+\frac{27}{8}s}g^4+\paren{-\frac{3}{8}+\frac{9}{4}s}g^2 \gp{2} + \paren{\frac{3}{2}+\frac{485}{48}s}\gp{4} \right. \nonumber\\
&+ \left. \paren{\frac{45}{4}g^2+\frac{85}{12}\gp{2}+40g_s^2}y_t^2 + \paren{18 - \frac{63}{2}s}y_t^4 + \paren{36g^2+12\gp{2}-36y_t^2}\paren{1+s}\lambda \right. \nonumber\\
&+ \left. \paren{-108+126s-144s^2+66s^3}\lambda^2 \frac{}{}\right].
\end{align}
In addition, the Higgs anomalous dimension $\gamma = -d\ln h / dt$ is given by
\begin{align}
\label{eq:gamma}
\gamma = &-\frac{1}{\paren{4\pi}^2}\left[ \frac{9}{4}g^2 + \frac{3}{4}\gp{2} - 3y_t^2 \right] \nonumber\\
&-\frac{1}{\paren{4\pi}^4}\left[ \frac{271}{32} g^4 - \frac{9}{16}g^2 \gp{2} - \frac{431}{96} s \gp{4} - \frac{5}{2}\paren{\frac{9}{4}g^2+\frac{17}{12}\gp{2}+8g_s^2}y_t^2+\frac{27}{4}sy_t^4-6s^3\lambda^2 \right].
\end{align}

The RG equations \eqref{eq:betalambda}--\eqref{eq:gamma} can easily be extended to include (i) the complete three-loop expressions for the gauge coupling beta functions~\cite{Mil12} and (ii) the leading three-loop corrections to $\beta_\lambda$, $\beta_{y_t}$, and $\gamma$~\cite{Che12}. These improvements can be made by adding the following terms to the beta functions:

\begin{align}
\Delta \beta_\lambda &= \frac{1}{\paren{4\pi}^6} \left[ \paren{7176+4032\zeta_3}\lambda^4 + 1746 y_t^2 \lambda^3 + \paren{1719 + 1512 \zeta_3}y_t^4 \lambda^2 + \paren{\frac{117}{4}-396 \zeta_3} y_t^6 \lambda \right. \nonumber\\
&\quad- \paren{\frac{1599}{4}+72 \zeta_3} y_t^8 + \paren{-2448+2304 \zeta_3} g_s^2 y_t^2 \lambda^2 + \paren{1790-2592 \zeta_3} g_s^2 y_t^4 \lambda  \nonumber \\
&\quad+ \left. \paren{-76 + 480 \zeta_3} g_s^2 y_t^6 + \paren{\frac{2488}{3} - 96 \zeta_3} g_s^4 y_t^2 \lambda + \paren{-\frac{532}{3} + 64 \zeta_3} g_s^4 y_t^4 \right], \label{eq:deltabetalambda}
\end{align}
\begin{align}
\Delta \beta_{y_t} &= \frac{y_t}{\paren{4\pi}^6} \left[ -36 \lambda^3 + \frac{15}{4} y_t^2 \lambda^2 + 198 y_t^4 \lambda +\paren{\frac{339}{8}+\frac{27}{2} \zeta_3} y_t^6 + 16 g_s^2 y_t^2 \lambda \right. \nonumber \\
&\quad- \left. 157 g_s^2 y_t^4 + \paren{\frac{3827}{6}-228 \zeta_3} g_s^4 y_t^2 + \paren{-\frac{4166}{3}+640 \zeta_3} g_s^6 \right],\label{eq:deltabetayt}
\end{align}
\begin{align}
\Delta \beta_{g^\prime} &= \frac{\gp{3}}{\paren{4\pi}^6}\left[\frac{1315}{64}g^4 + \frac{205}{96}g^2 \gp{2} - \frac{388613}{5184}\gp{4} - g^2 g_s^2 - \frac{137}{27} \gp{2} g_s^2 + 99g_s^4 \right. \nonumber \\ 
&\quad- \left. y_t^2\paren{\frac{785}{32}g^2 + \frac{2827}{288}\gp{2} + \frac{29}{3} g_s^2} + \frac{315}{16} y_t^4 + \lambda \paren{\frac{3}{2}g^2 + \frac{3}{2}\gp{2} - 3\lambda}\right], \label{eq:deltabetagp}
\end{align}
\begin{align}
\Delta \beta_g &= \frac{g^3}{\paren{4\pi}^6}\left[\frac{324953}{1728}g^4 + \frac{291}{32}g^2 \gp{2} - \frac{5597}{576}\gp{4} + 39g^2 g_s^2 - \frac{1}{3} \gp{2} g_s^2 + 81g_s^4 \right. \nonumber \\ 
&\quad- \left. y_t^2\paren{\frac{729}{32}g^2 + \frac{593}{96}\gp{2} + 7g_s^2} + \frac{147}{16} y_t^4 + \lambda \paren{\frac{3}{2}g^2 + \frac{1}{2}\gp{2} - 3\lambda}\right], \label{eq:deltabetag}
\end{align}
\begin{align}
\Delta \beta_{g_s} &= \frac{g_s^3}{\paren{4\pi}^6}\left[\frac{109}{8}g^4 - \frac{1}{8}g^2 \gp{2} - \frac{2615}{216}\gp{4} + 21g^2 g_s^2 + \frac{77}{9} \gp{2} g_s^2 + \frac{65}{2}g_s^4 \right. \nonumber \\ 
&\quad- \left. y_t^2\paren{\frac{93}{8}g^2 + \frac{101}{24}\gp{2} + 40g_s^2} + 15 y_t^4 \right], \label{eq:deltabetags}
\end{align}
\begin{align}
\Delta \gamma &= -\frac{1}{\paren{4\pi}^6} \left[ 36 \lambda^3 + \frac{135}{2} y_t^2 \lambda^2 - 45 y_t^4 \lambda - \paren{\frac{789}{16}+9 \zeta_3} y_t^6 \right. \nonumber \\
&\quad- \left. \paren{\frac{15}{2} - 72 \zeta_3} g_s^2 y_t^4 - \paren{\frac{622}{3}-24 \zeta_3} g_s^4 y_t^2 \right], \label{eq:deltagamma}
\end{align}
where $\zeta_3 \equiv \zeta(3) \simeq 1.202$.


\begin{thebibliography}{99}

\bibitem{Sta80}
A.~Starobinsky, {\it A new type of isotropic cosmological models without singularity},
{\em Phys. Lett. B} {\bf 91} (1980) 99--102

\bibitem{Gut81}
A.~Guth, {\it Inflationary universe: A possible solution to the horizon and flatness problems},
{\em Phys. Rev. D} {\bf 23} (1981) 347--356

\bibitem{Lin82}
A.~Linde, {\it A new inflationary universe scenario: A possible solution of the horizon, flatness, homogeneity, isotropy and primordial monopole problems},
{\em Phys. Lett. B} {\bf 108} (1982) 389--393

\bibitem{Alb82}
A.~Albrecht and P.J.~Steinhardt, {\it Cosmology for Grand Unified Theories with radiatively induced symmetry breaking},
{\em Phys. Rev. Lett.} {\bf 48} (1982) 1220--1223

\bibitem{Lin83}
A.~Linde, {\it Chaotic inflation},
{\em Phys. Lett. B} {\bf 129} (1983) 177--181

\bibitem{WMA12}
WMAP~collaboration, {\it Nine-year Wilkinson Microwave Anisotropy Probe (WMAP) observations: cosmological parameter results},
[\href{http://arxiv.org/abs/1212.5226}{\tt arXiv:1212.5226}]

\bibitem{Pla13}
PLANCK~collaboration, {\it Planck 2013 results. XXII. Constraints on inflation},
[\href{http://arxiv.org/abs/1303.5082}{{\tt arXiv:1303.5082}}]

\bibitem{Lid93}
A.~Liddle and D.~Lyth, {\it The cold dark matter density perturbation},
{\em Phys. Rept.} {\bf 231} (1993) 1--105,
[\href{http://arxiv.org/abs/astro-ph/9303019}{\tt astro-ph/9303019}]

\bibitem{Lyt99}
D.~Lyth and A.~Riotto, {\it Particle physics models of inflation and the cosmological density perturbation},
{\em Phys. Rept.} {\bf 314} (1999) 1--146,
[\href{http://arxiv.org/abs/hep-ph/9807278}{\tt hep-ph/9807278}]

\bibitem{Mar13}
J~Martin, C~Ringeval and V~Vennin, {\it Encyclopaedia Inflationaris},
[\href{http://arxiv.org/abs/arXiv:1303.3787}{\tt arXiv:1303.3787}]

\bibitem{Bez08}
F.~Bezrukov and M.~Shaposhnikov, {\it The standard model Higgs boson as the
  inflaton},  {\em Phys. Lett. B} {\bf 659} (2008) 703--706,
  [\href{http://arxiv.org/abs/0710.3755}{\tt arXiv:0710.3755}].

\bibitem{Isi08}
G.~Isidori, V.~Rychkov, A.~Strumia and N.~Tetradis, {\it Gravitational corrections to the
  standard model vacuum decay},  {\em Phys. Rev. D} {\bf 77} (2008) 025034,
  [\href{http://arxiv.org/abs/0712.0242}{\tt arXiv:0712.0242}].

\bibitem{Mas12a}
I.~Masina and A.~Notari, {\em Standard Model false vacuum inflation: Correlating the tensor-to-scalar ratio to the top quark and Higgs boson masses},
{\em Phys. Rev. Lett.} {\bf 108} (2012) 191302,
[\href{http://arxiv.org/abs/arXiv:1112.5430}{\tt arXiv:1112.5430}].

\bibitem{Mas12b}
I.~Masina and A.~Notari, {\em The Higgs mass range from standard model false vacuum inflation in scalar-tensor gravity},
{\em Phys. Rev. D} {\bf 85} (2012) 123506,
[\href{http://arxiv.org/abs/1112.2659}{\tt arXiv:1112.2659}].

\bibitem{Mas12c}
I.~Masina and A.~Notari, {\em Inflation from the Higgs field false vacuum with hybrid potential},
{\em JCAP} {\bf 11} (2012) 031,
[\href{http://arxiv.org/abs/1204.4155}{\tt arXiv:1204.4155}].

\bibitem{Ger10}
C.~Germani and A.~Kehagias, {\em New Model of Inflation with Nonminimal Derivative Coupling of Standard Model Higgs Boson to Gravity}, {\em Phys. Rev. Lett.} {\bf 105} (2010) 011302,
[\href{http://arxiv.org/abs/arXiv:1003.2635}{\tt arXiv:1003.2635}].

\bibitem{Nak10}
K.~Nakayama and F.~Takahashi, {\it Running kinetic inflation}, {\em JCAP} {\bf 11} (2010) 009,
[\href{http://arxiv.org/abs/arXiv:1008.2956}{\tt arXiv:1008.2956}].

\bibitem{Kam11}
K.~Kamada, T.~Kobayashi, M.~Yamaguchi and J.~Yokoyama, {\it Higgs $G$ inflation}, {\em Phys. Rev. D} {\bf 83} (2011) 083515,
[\href{http://arxiv.org/abs/arXiv:1012.4238}{\tt arXiv:1012.4238}].

\bibitem{Kam12}
K.~Kamada, T.~Kobayashi, T.~Takahashi, M.~Yamaguchi and J.~Yokoyama, {\it Generalized Higgs inflation}, {\em Phys. Rev. D} {\bf 86} (2012) 023504,
[\href{http://arxiv.org/abs/arXiv:1203.4059}{\tt arXiv:1203.4059}].

\bibitem{Her12}
M.~Hertzberg, {\it Can inflation be connected to low energy particle physics?}, {\em JCAP} {\bf 08} (2012) 008,
[\href{http://arxiv.org/abs/arXiv:1110.5650}{\tt arXiv:1110.5650}].

\bibitem{Sal89}
D.~Salopek, J.~Bond and J.~Bardeen, {\it Designing density fluctuation spectra in
  inflaton},  {\em Phys. Rev. D} {\bf 40} (1989) 1753--1788,
  [\href{http://inspirehep.net/record/265815}{\tt INSPIRE-HEP}].
  
\bibitem{Fak90}
R.~Fakir and W.~Unruh, {\it Improvement on cosmological chaotic inflation through nonminimal coupling},
{\em Phys. Rev. D} {\bf 41} (1990) 1783--1791

\bibitem{Kai95}
D.~Kaiser, {\it Primordial spectral indices from generalized Einstein theories},
{\em Phys. Rev. D} {\bf 52} (1995) 4295--4306,
[\href{http://arxiv.org/abs/astro-ph/9408044}{\tt astro-ph/9408044}].

\bibitem{Kom99}
E.~Komatsu and T.~Futamase, {\it Complete constraints on a nonminimally coupled chaotic inflationary scenario from the cosmic microwave background}, {\em Phys.Rev. D} {\bf 59} (1999) 064029
[\href{http://arxiv.org/abs/astro-ph/9901127}{\tt astro-ph/9901127}].

\bibitem{Gia13}
P.~Giardino, K.~Kannike, I.~Masina, M.~Raidal and A.~Strumia, {\it The universal Higgs fit},
[\href{http://arxiv.org/abs/1303.3570}{\tt arXiv:1303.3570}].

\bibitem{Bur09}
C.~Burgess, H.~Lee and M.~Trott, {\it Power-counting and the validity of the classical approximation during inflation}, {\em JHEP} {\bf 09} (2009) 103,
[\href{http://arxiv.org/abs/0902.4465}{\tt arXiv:0902.4465}].

\bibitem{Bar09}
J.~Barbon and J.~Espinosa, {\it On the naturalness of Higgs inflation}, {\em Phys. Rev. D.} {\bf 79} (2009) 081302(R),
[\href{http://arxiv.org/abs/0903.0355}{\tt arXiv:0903.0355}].

\bibitem{Bur10}
C.~Burgess, H.~Lee and M.~Trott, {\it On Higgs inflation and naturalness}, {\em JHEP} {\bf 07} (2007) 007,
[\href{http://arxiv.org/abs/1002.2730}{\tt arXiv:1002.2730}].

\bibitem{Her10}
M.~Hertzberg, {\it On inflation with non-minimal coupling}, {\em JHEP} {\bf 11} (2010) 023,
[\href{http://arxiv.org/abs/1002.2995}{\tt arXiv:1002.2995}].

\bibitem{Bez09b}
F.~Bezrukov and M.~Shaposhnikov, {\it Standard Model Higgs boson mass from inflation: two loop analysis},
  {\em JCAP} {\bf 07} (2009) 089,
  [\href{http://arxiv.org/abs/arXiv:0904.1537}{\tt arXiv:0904.1537}].

\bibitem{Bez11}
F.~Bezrukov, A.~Magnin, M.~Shaposhnikov and S.~Sibiryakov, {\it Higgs inflation: consistency and generalisations},  {\em JHEP} {\bf 01} (2011) 016,
[\href{http://arxiv.org/abs/1008.5157}{\tt arXiv:1008.5157}].

\bibitem{Fer11}
S.~Ferrara, R.~Kallosh, A.~Linde, A.~Marrani and A.~Van~Proeyen, {\it Superconformal symmetry, NMSSM, and inflation}, {\em Phys. Rev. D} {\bf 83} (2011) 025008,
[\href{http://arxiv.org/abs/1008.2942}{\tt arXiv:1008.2942}].

\bibitem{Ler12}
R.~Lerner and J.~McDonald, {\it Unitarity-violation in generalized Higgs inflation models},
{\em JCAP} {\bf 11} (2012) 019,
[\href{http://arxiv.org/abs/1112.0954}{\tt arXiv:1112.0954}].
  
\bibitem{Ler10}
R.~Lerner and J.~McDonald, {\it A unitarity-conserving Higgs inflation model},
{\em Phys. Rev. D} {\bf 82} (2010) 103525,
[\href{http://arxiv.org/abs/1005.2978}{\tt arXiv:1005.2978}].

\bibitem{DeS09}
A.~De Simone, M.~Hertzberg and F.~Wilczek, {\it Running inflation in the Standard Model},
  {\em Phys. Lett. B} {\bf 678} (2009) 1--8,
  [\href{http://arxiv.org/abs/0812.4946}{\tt arXiv:0812.4946}].

\bibitem{Bez11b}
F.~Bezrukov, D.~Gorbunov and M.~Shaposhnikov, {\it Late and early time phenomenology of Higgs-dependent cutoff}, {\em JCAP} {\bf 10} (2011) 001,
[\href{http://arxiv.org/abs/1106.5019}{\tt arXiv:1106.5019}].

\bibitem{Bez12}
F.~Bezrukov, M.~Kalmykov, B.~Kniehl and M.~Shaposhnikov, {\it Higgs boson mass and new physics},
{\em JHEP} {\bf 10} (2012) 140,
[\href{http://arxiv.org/abs/1205.2893}{\tt arXiv:1205.2893}].

\bibitem{Deg12}
G.~Degrassi, S.~Di~Vita, J.~Elias-Miro, J.~Espinosa, G.~Giudice, G.~Isidori and A. Strumia, {\it Higgs mass and vacuum stability in the Standard Model at NNLO}, {\em JHEP} {\bf 08} (2012) 098, [\href{http://arxiv.org/abs/1205.6497}{\tt arXiv:1205.6497}].

\bibitem{Sha10}
M.~Shaposhnikov and C.~Wetterich, {\it Asymptotic safety of gravity and the Higgs boson mass},
{\em Phys. Lett. B} {\bf 683} (2010) 192--200,
[\href{http://arxiv.org/abs/0912.0208}{\tt arXiv:0912.0208}].

\bibitem{Bez09a}
F.~Bezrukov, A.~Magnin and M.~Shaposhnikov, {\it Standard Model Higgs boson mass from inflation},
  {\em Phys. Lett. B} {\bf 675} (2008) 88--92,
  [\href{http://arxiv.org/abs/arXiv:0812.4950}{\tt arXiv:0812.4950}].

\bibitem{Cla09}
T.~Clark, B.~Liu, S.~Love and T.~ter~Veldhuis, {\it Standard model Higgs boson-inflaton and dark matter}, {\em Phys. Rev. D} {\bf 80} (2009) 075019,
[\href{http://arxiv.org/abs/0906.5595}{\tt arXiv:0906.5595}].

\bibitem{Ler09}
R.~Lerner and J.~McDonald, {\it Gauge singlet scalar as inflaton and thermal relic dark matter},
{\em Phys. Rev. D} {\bf 80} (2009) 123507,
[\href{http://arxiv.org/abs/0909.0520}{\tt arXiv:0909.0520}].

\bibitem{Ler11}
R.~Lerner and J.~McDonald, {\it Distinguishing Higgs inflation and its variants},
{\em Phys. Rev. D} {\bf 83} (2011) 123522,
[\href{http://arxiv.org/abs/1104.2468}{\tt arXiv:1104.2468}].

\bibitem{Mil12}
L.~Milhaila, J.~Salomon and M.~Steinhauser, {\it Gauge coupling $\beta$-functions in the standard model to three loops}, {\em Phys. Rev. Lett.} {\bf 108} (2012) 151602,
[\href{http://arxiv.org/abs/1201.5868}{\tt arXiv:1201.5868}].

\bibitem{Che12}
K.~Chetyrkin and M.~Zoller, {\it Three-loop $\beta$-function for top-Yukawa and the Higgs self-interaction in the standard model}, {\em JHEP} {\bf 06} (2012) 033,
[\href{http://arxiv.org/abs/1205.2892}{\tt arXiv:1205.2892}].

\bibitem{Bez08b}
F.~Bezrukov, {\it The Standard model Higgs as the inflaton},
[\href{http://arxiv.org/abs/0805.2236}{\tt arXiv:0805.2236}].

\bibitem{Bez09c}
F.~Bezrukov, D.~Gorbunov and M.~Shaposhnikov, {\it On initial conditions for the hot big bang}, {\em JCAP} {\bf 06} (2009) 029,
[\href{http://arxiv.org/abs/0812.3622}{\tt arXiv:0812.3622}].

\bibitem{Gar09}
J.~Garcia-Bellido, D.~Figueroa, and J.~Rubio, {\it Preheating in the standard model with the Higgs inflaton coupled to gravity}, {\em Phys. Rev. D} {\bf 79} (2009) 063531,
[\href{http://arxiv.org/abs/0812.4624}{\tt arXiv:0812.4624}].

\bibitem{Dut08}
S.~Dutta, K.~Hagiwara, Q.-S.~Yan and K.~Yoshida, {\it Constraints on the electroweak chiral Lagrangian from the precision data}, {\em Nucl. Phys. B} {\bf 790} (2008) 111,
[\href{http://arxiv.org/abs/0705.2277}{\tt arXiv:0705.2277}].

\bibitem{Buc95}
W.~Buchmuller and O.~Philipsen, {\it Phase structure and phase transition of the SU(2) Higgs model in three dimensions}, {\em Nucl. Phys. B} {\bf 443} (1995) 47--69,
[\href{http://arxiv.org/abs/hep-ph/9411334}{\tt hep-ph/9411334}].

\bibitem{Ara92}
H.~Arason, D.~Castano, B.~Kesthelyi, S.~Mikaelian, E.~Piard, P.~Ramond and B.~Wright, {\it Renormalization-group study of the standard model and its extensions: The standard model}, {\em Phys. Rev. D} {\bf 46} (1992) 3945--3965

\bibitem{Baa13}
M.~Baak and R.~Kogler, {\it The global electroweak Standard Model fit after the Higgs discovery},
[\href{http://arxiv.org/abs/1306.0571}{\tt arXiv:1306.0571}].

\bibitem{For92}
C.~Ford, I.~Jack and D.~Jones, {\it The standard model effective potential at two loops}, {\em Nucl. Phys. B} {\bf 387} (1992) 373--390,
[\href{http://arxiv.org/abs/hep-ph/0111190}{\tt hep-ph/0111190}].

\bibitem{Bar08}
A.~Barvinsky, A.~Kamenshchik and A.~Starobinsky, {\it Inﬂation scenario via the Standard Model Higgs boson and LHC}, {\em JCAP} {\bf 11} (2008) 021,
[\href{http://arxiv.org/abs/0809.2104}{\tt arXiv:0809.2104}].
  
\bibitem{Sha09a}
M.~Shaposhnikov and D.~Zenhausern, {\it Scale invariance, unimodular gravity and dark energy}, {\em Phys. Lett. B} {\bf 671} (2009) 187--192,
[\href{http://arxiv.org/abs/0809.3395}{\tt arXiv:0809.3395}].
  
\bibitem{Sha09b}
M.~Shaposhnikov and D.~Zenhausern, {\it Quantum scale invariance, cosmological constant and hierarchy problem}, {\em Phys. Lett. B} {\bf 671} (2009) 162--168,
[\href{http://arxiv.org/abs/0809.3406}{\tt arXiv:0809.3406}].

\bibitem{Bez13}
F.~Bezrukov, G.~Karananas, J.~Rubio and M.~Shaposhnikov, {\it Higgs-dilaton cosmology: An effective field theory approach}, {\em Phys. Rev. D} {\bf 87} (2013) 096001,
[\href{http://arxiv.org/abs/1212.4148}{\tt arXiv:1212.4148}].

\bibitem{Moo11}
S.~Mooij and M.~Postma, {\it Goldstone bosons and a dynamical Higgs field}, {\em JCAP} {\bf 09} (2011) 006,
[\href{http://arxiv.org/abs/1104.4897}{\tt arXiv:1104.4897}].

\bibitem{Geo12}
D.~George, S.~Mooij and M.~Postma, {\it Effective action for the Abelian Higgs model in FLRW}, {\em JCAP} {\bf 11} (2012) 043,
[\href{http://arxiv.org/abs/1207.6963}{\tt arXiv:1207.6963}].

\bibitem{Sal13}
A.~Salvio, {\it Higgs inflation at NNLO after the boson discovery},
[\href{http://arxiv.org/abs/1308.2244}{\tt arXiv:1308.2244}].

\bibitem{For93}
C.~Ford, D.~Jones and P.~Stephenson, {\it The effective potential and the renormalization group},
{\em Nucl. Phys. B} {\bf 395} (1993) 17--34,
[\href{http://arxiv.org/abs/hep-lat/9210033}{\tt hep-lat/9210033}].

\bibitem{Bur10b}
C.~Burigana, C.~Destri, H.~J.~de~Vega, A.~Gruppuso, N.~Mandolesi, P.~Natoli and N.~G.~Sanchez, {\it Forecast for the Planck Precision on the tensor-to-scalar ratio and other cosmological parameters}, {\em Astrophys. J.} {\bf 724} (2010) 588--607,
[\href{http://arxiv.org/abs/1003.6108}{\tt arXiv:1003.6108}].

\bibitem{Asa05}
T.~Asaka, S.~Blanchet and M.~Shaposhnikov, {\it The $\nu$MSM, dark matter and
  neutrino masses},  {\em Phys. Lett. B} {\bf 631} (2005) 151--156,
  [\href{http://arxiv.org/abs/hep-ph/0503065}{{\tt hep-ph/0503065}}].

\bibitem{Sha05}
T.~Asaka and M.~Shaposhnikov, {\it The $\nu$MSM, dark matter and baryon
  asymmetry of the universe},  {\em Phys. Lett. B} {\bf 620} (2005) 17--26,
  [\href{http://arxiv.org/abs/hep-ph/0505013}{{\tt hep-ph/0505013}}].

\end{thebibliography}
\end{document}